\documentclass[trackchanges,twocolumn]{aastex631}

\usepackage{graphicx} 
\usepackage{multirow}
\usepackage{xcolor}
\usepackage{makecell}
\usepackage{xfrac}

\sfcode`A=1000 \sfcode`B=1000 \sfcode`C=1000 \sfcode`D=1000
\sfcode`E=1000 \sfcode`F=1000 \sfcode`G=1000 \sfcode`H=1000
\sfcode`I=1000 \sfcode`J=1000 \sfcode`K=1000 \sfcode`L=1000
\sfcode`M=1000 \sfcode`N=1000 \sfcode`O=1000 \sfcode`P=1000
\sfcode`Q=1000 \sfcode`R=1000 \sfcode`S=1000 \sfcode`T=1000
\sfcode`U=1000 \sfcode`V=1000 \sfcode`W=1000 \sfcode`X=1000
\sfcode`Y=1000 \sfcode`Z=1000

\def\zA{$z_A = 0.6850$}
\def\zAErr{$z_A = 0.68501 \pm 0.00015$}
\def\zB{$z_B = 0.6835$}
\def\zBErr{$z_B = 0.68348 \pm 0.00008$}

\def\zSMGB{$z_{{\rm SMG}} = 2.221$}
\def\zSMGBErr{$z_{{\rm SMG}} = 2.2207 \pm 0.0012$}

\newcommand{\zSMG}{\zSMGB}

\newcommand{\js}{}
\def\chandra {{\js Chandra}}
\def\hst {{\js HST}}

\def\spitzer {{\js Spitzer}}
\def\ciao {\textsc{ciao}}
\def\dsn {\textsc{ds9}}
\def\Bagpipes {\textsc{Bagpipes}}
\def\RCD {\hbox{$R_{\rm CD}$}}
\def\Bint {$B_{{\rm int}}$}
\def\BintM {B_{{\rm int}}}
\def\Bme {$B_{{\rm me}}$}
\def\BmeM {B_{{\rm me}}}
\def\kms{km~s$^{-1}$}

\def\Oii {\hbox{[\ion{O}{2}]}}
\def\Oiii {\hbox{[\ion{O}{3}]}}
\def\Mgii {\hbox{\ion{Mg}{2}}}
\def\Cii {\hbox{\ion{C}{2}]}}
\def\Hbeta {\hbox{H$\beta$}}
\newcommand{\Lsun}{\hbox{L$_\sun$}}
\newcommand{\Msun}{\hbox{M$_\sun$}}
\newcommand{\BAYMAX}{{\tt BAYMAX}}

\newcommand{\uJy}{$\mu$Jy}
\newcommand{\no}{\nodata}
\newcommand{\0}{\phantom{0}}
\defcitealias{Haas_et_al}{H14} 	

\received{February 15, 2023}
\revised{July 23, 2024}
\accepted{July 25, 2024}
\submitjournal{ApJ}

\shorttitle{3C~220.3 Lensed System}
\shortauthors{Hyman et al.}

\begin{document}

\title{A Multiwavelength Portrait of the 3C~220.3 Lensed System}

\correspondingauthor{S\'oley Hyman}
\email{soleyhyman@arizona.edu}

\author[0000-0002-6036-1858]{S\'oley \'O.\ Hyman}
\affiliation{Center for Astrophysics \textbar\ Harvard \& Smithsonian, 60 Garden St., Cambridge, MA 02138, USA}
\affiliation{Univ.\ of Arizona, Dept.\ of Astronomy, Tucson, AZ 85719, USA}
\author[0000-0003-1809-2364]{Belinda J.\ Wilkes}
\affiliation{Center for Astrophysics \textbar\ Harvard \& Smithsonian, 60 Garden St., Cambridge, MA 02138, USA}
\affiliation{H.\ H.\ Wills Physics Laboratory, Univ.\ of Bristol, Bristol BS8 1TL, UK}
\author[0000-0002-9895-5758]{S.~P.~Willner}
\affiliation{Center for Astrophysics \textbar\ Harvard \& Smithsonian, 60 Garden St., Cambridge, MA 02138, USA}
\author[0000-0001-5513-029X]{Joanna Kuraszkiewicz}
\affiliation{Center for Astrophysics \textbar\ Harvard \& Smithsonian, 60 Garden St., Cambridge, MA 02138, USA}
\author[0000-0001-6004-9728]{Mojegan Azadi}
\affiliation{Center for Astrophysics \textbar\ Harvard \& Smithsonian, 60 Garden St., Cambridge, MA 02138, USA}
\author[0000-0002-1516-0336]{D.~M.~Worrall}
\affiliation{H.\ H.\ Wills Physics Laboratory, Univ.\ of Bristol, Bristol BS8 1TL, UK}
\author[0000-0002-1616-1701]{Adi Foord}
\affiliation{Department of Physics, University of Maryland Baltimore County, 1000 Hilltop Cir, Baltimore, MD 21250, USA}
\author{Simona Vegetti}
\affiliation{Max-Planck-Institut f\"ur Astrophysik, Garching, Germany}
\author[0000-0002-3993-0745]{Matthew L.~N.\ Ashby}
\affiliation{Center for Astrophysics \textbar\ Harvard \& Smithsonian, 60 Garden St., Cambridge, MA 02138, USA}
\author[0000-0002-1858-277X]{Mark Birkinshaw}
\altaffiliation{Author is deceased}
\affiliation{H.\ H.\ Wills Physics Laboratory, Univ.\ of Bristol, Bristol BS8 1TL, UK}
\author[0000-0002-4030-5461]{Christopher Fassnacht}
\affiliation{University of California, Davis, CA 95616, USA}
\author[0000-0002-7284-0477]{Martin Haas}
\affiliation{Astronomisches Institut, Ruhr Universit\"at, Bochum, Germany}
\author[0000-0003-2686-9241]{Daniel Stern}
\affiliation{Jet Propulsion Laboratory, California Institute of Technology, Pasadena, CA 91109, USA}

\begin{abstract}
The 3C~220.3 system is a rare case of a foreground narrow-line radio galaxy  (``galaxy A," \zA) lensing a background submillimeter galaxy (\zSMG). New spectra from MMT/Binospec confirm that the companion galaxy (``galaxy B") is part of the lensing system with \zB.  New three-color \hst\ data reveal a full Einstein ring and allow a more precise lens model. The new HST images also reveal extended emission around galaxy A, and the spectra show extended \Oii\ emission with irregular morphology and complex velocity structure.  All indications are that the two lensing galaxies are a gravitationally interacting pair. Strong \Oii\ emission from both galaxies A and B suggests current star formation, which could be a consequence of the interaction. This would indicate a younger stellar population than previously assumed and imply smaller stellar masses for the same luminosity. The improved lens model and expanded spectral energy distributions have enabled better stellar mass estimates for the foreground galaxies. The resulting dark matter fractions are $\sim$0.8, which are higher than previously calculated. Deeper \chandra\ imaging shows extended X-ray emission but no evidence for an X-ray point source associated with either galaxy. The detection of X-rays from the radio lobes of 3C~220.3 allows an estimate of $\sim$3~nT for the magnetic fields in the lobes, a factor of $\sim$3 below the equipartition fields, as is typical for radio galaxies.
\end{abstract}


\section{Introduction} \label{sec:intro}

In a 2012 survey of low- to mid-redshift radio galaxies and quasars \citep{Westhues_2016}, the narrow-line radio galaxy (NLRG) \object{3C~220.3} (Fanaroff--Riley type II, redshift $z=0.685$) was discovered to have unusually strong far-infrared (FIR) emission for its classification and redshift \citep[referred to hereafter as H14]{Haas_et_al}---over 10~times greater than similarly selected galaxies.\footnote{3C~220.3 also stood out in an earlier study by \citet{3C220p3_L_178MHz_Cleary_2007}. In an analysis of \spitzer\ data for 33 3CRR sources, it was the only target to have a detection at 160~\micron. However, the galaxy was not investigated further at that time.} An archival Hubble Space Telescope (\hst) image showed an arc of emission suggestive of a gravitationally lensed background source (Figure~\ref{fig:new-hst-einstein-rings}). Follow-up with Keck NIRC2 and the Submillimeter Array (SMA) revealed a close-to-full Einstein ring with a radius of 1\farcs8, confirming the presence of a lensed background source. Visible light spectroscopy revealed the source of the high FIR flux to be a rare submillimeter galaxy (SMG), designated as \object{SMM~J0939+8315} (referred to hereafter as ``the SMG") at $z=2.221$ \citepalias{Haas_et_al}. In the X-ray regime, 3C~220.3 is a low-count, diffuse source \citepalias{Haas_et_al} and the active galactic nucleus (AGN) (galaxy A) appears to be Compton thick \citep{Kuras2021}, consistent with its weak radio core and NLRG classification. 

3C~220.3 shows no evidence of association with a massive galaxy cluster \citepalias{Haas_et_al}. As such, the primary deflector of gravitational lensing is the radio galaxy host (i.e., galaxy~A, \citetalias{Haas_et_al}), but the archival \hst\ image also revealed a second object (galaxy~B) located 1\farcs5 south of galaxy~A, within the radius of the lens arc. 
Optical spectroscopy confirmed galaxy~A's $z = 0.685$ and determined the SMG to be at $z=2.221$. Although the redshift of galaxy~B was unknown, the lens modeling suggested that it is close to that of galaxy~A \citepalias{Haas_et_al}.
A 9~GHz radio image from the NSF's Karl G.\ Jansky Very Large Array (VLA) revealed an unresolved core in addition to the known lobe emission (\citetalias[their Fig.~1]{Haas_et_al}; \citealp[Fig.~8]{hyman_thesis}). A 10~ks observation by \chandra\ showed diffuse emission extending over the full system, but with only 16~counts, it was impossible to associate the X-rays with individual components of the system.

The 3C~220.3 system is complex. It includes radio lobes and a faint radio core at the host-galaxy location, spatially extended visible emission from the host galaxy~A, a visible companion galaxy~B, diffuse X-ray emission across the entire system, and the gravitationally lensed SMG. The system is a prime candidate for  multi-wavelength study, facilitating constraints on  magnetic field strengths, dark matter fractions, the nature of the two lensing galaxies, and the opportunity to study a rare, lensed, high-redshift SMG.

This paper extends \citetalias{Haas_et_al}'s earlier analysis of 3C~220.3 with deeper X-ray data from \chandra, visible light imaging from \hst, and MMT optical long-slit spectra. These observations enable more accurate mass estimates of the 3C~220.3 system (useful in both dark matter calculations and lens modeling) and better trace the origin(s) of the diffuse X-ray emission that is distributed throughout the system. New optical spectra from the MMT telescope also determine the redshift of galaxy~B.

This paper is structured as follows: Section~\ref{sec:obs-data} provides an overview of the data used in the paper. Section~\ref{sec:analysis-results} discusses the analysis of the various 3C~220.3 system components through visible--near-infrared (NIR) imaging and spectroscopy, as well as X-ray imaging and modeling. Section~\ref{sec:discussion} discusses the properties of the galaxy system, specifically spectral energy distributions (SEDs), stellar mass calculations of galaxies A and B, the SMG lens modeling, and magnetic field strengths. We adopt a flat $\Lambda$CDM cosmological model with 
$H_0 = 67.37$~km~s$^{-1}$~Mpc$^{-1}$, $\Omega_{\Lambda} = 0.6854$, and $\Omega_M = 0.3147$ \citep{Planck2018_cosmology}.

\section{Observations} 
\label{sec:obs-data}
\subsection{Hubble Space Telescope}
\label{subsec:hst-data}

Two archival \hst/Planetary Camera exposures were taken with the F702W filter (702~nm) on 1994 March~1 and 1995 May~5 for 300~seconds each. The resulting images  \citepalias[Fig.~1]{Haas_et_al}\footnote{Sparks, William, 1994, Hubble Space Telescope Cycle~4 Proposal \#5476, ``Continuum Snapshots of 3CR Radio Galaxies.''\label{fn:hst-prop-94}} show the arc of a partial Einstein ring to the northeast of the radio galaxy core. The arc is approximately 2\arcsec\ in radius and subtends $\sim$65\degr\ centered near galaxy~A. 

\begin{figure*}[hbt!]
\includegraphics[width=\textwidth,trim={3cm 1.7cm 2.4cm 1.6cm},clip]{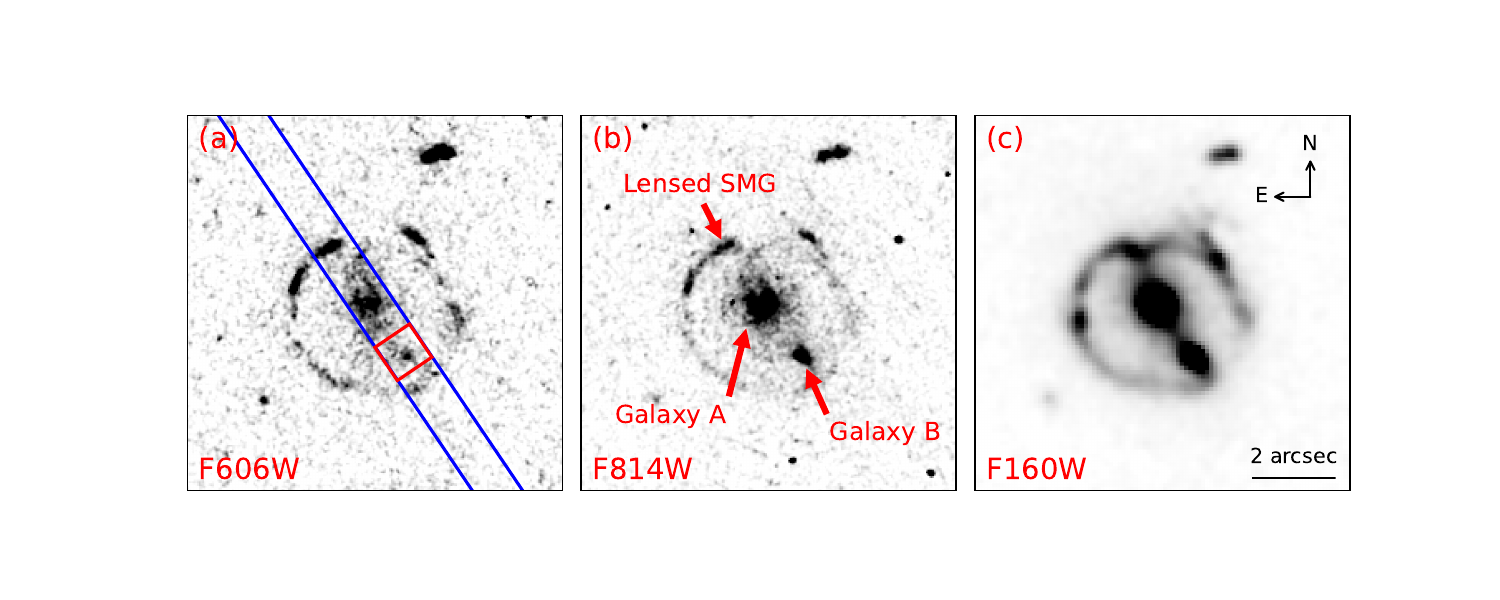}

\caption{Smoothed (Gaussian with $r=1.5$~pix) \hst\ negative images of the 3C~220.3 system. Images (left to right) have effective wavelengths of 606~nm, 814~nm, and 1.60~\micron. Images are 9\arcsec\ on a side with north up and east to the left. System components are labeled in the center panel. The bright spot to the left of the galaxy~A core in F814W is a cosmic ray which could not be removed during image drizzling (C.~Leipski, private comm.\ 2014). The 2021 Binospec slit (20\arcsec$\times$1\arcsec) placement is shown in blue in panel (a) with the red box (1\arcsec$\times$1\arcsec) marking the center of the slit.}
\label{fig:new-hst-einstein-rings}
\end{figure*}

We obtained deeper \hst\ images on 2014 January 24 with the UVIS (F606W, F814W) and IR (F160W) detectors of the Wide Field Camera~3 (WFC3). Two dithered exposures were taken for each filter. The drizzling and cosmic ray rejection (Christian Leipski, private comm., 2014) used AstroDrizzle v.1.1.8, Numpy v.1.8.0, and PyFITS v.3.1.3.dev. Table~\ref{table:hst_data} lists exposure times, and Figure~\ref{fig:new-hst-einstein-rings} shows the resulting reduced images. All three show a full Einstein ring with a radius of 2\farcs0. The high-resolution F606W and F814W images also show that galaxy~B is located well inside the radius of the ring and is not part of the lensed SMG.\par

The 2014 \hst\ data place the AGN core (galaxy~A) at an offset from the VLA radio position. The VLA positions are more accurate than those of \hst\ (positional uncertainties $\lesssim$0\farcs1 for VLA\footnote{\url{https://science.nrao.edu/facilities/vla/docs/manuals/oss2012B/performance/positional-accuracy}} vs.\ ${\sim}1\arcsec$ for \hst; Section~3.1 of \citealt{2008MNRAS.388..421M}), and because the VLA image was used to align the other multiwavelength imaging, we shifted the \hst\ images to match galaxy~A to the VLA core. The shifts for each image were obtained by measuring centroids of the core in the VLA image and the three optical images using \dsn. Pre- and post-alignment images are shown in Figure~9 of \citet{hyman_thesis}, and Table~\ref{table:hst_data} gives the applied shifts.

\begin{table}[hbt!]
\caption{2014 \hst\ exposure parameters}
\label{table:hst_data}
\centering
\begin{tabular}{lchhhccc}
\hline \hline
Filter & Wavelength & Instrument & Date && Exposure & R.A. shift & Decl. shift \\
& (nm) &&&&time (s) & (\arcsec) & (\arcsec)\\
\hline
F606W & \0606 & WFC3/UVIS & 2014-01-24 & 2 \times 348.0 =& \0696 & $-$0.80 & +0.19 \\
F814W & \0814 & WFC3/UVIS & 2014-01-24 & 2 \times \0669.0 =& 1338 & $-$0.55 & +0.18 \\
F160W & 1600 & WFC3/IR & 2014-01-24 & 2 \times 302.9 =& \0606 & $-$1.36 & +0.13 \\
\hline
\end{tabular}
\raggedright
\tablecomments{Proposal \#13506, PI Wilkes, 2014-01-24. The tabulated R.A./Decl. shifts were applied after the initial data reduction to align galaxy~A with the more accurate VLA position of the radio core.}
\end{table}

\subsection{Chandra X-Ray Observations}
\label{subsec:chandra-data}

The first \chandra\ Advanced CCD Imaging Spectrometer \citep[ACIS-S;][]{ACIS_paper} observation of 3C~220.3 was taken on 2013 January 21 for 10~ks. The data showed 16 X-ray photons (with ${\sim}1.8$  expected from background) distributed across the $\sim$11\arcsec\ extent of the radio lobes \citepalias{Haas_et_al}. 

\begin{figure}[hbt]
\centering
\includegraphics[width=0.45\textwidth,trim={3.5cm 1.25cm 3cm 1.25cm},clip]{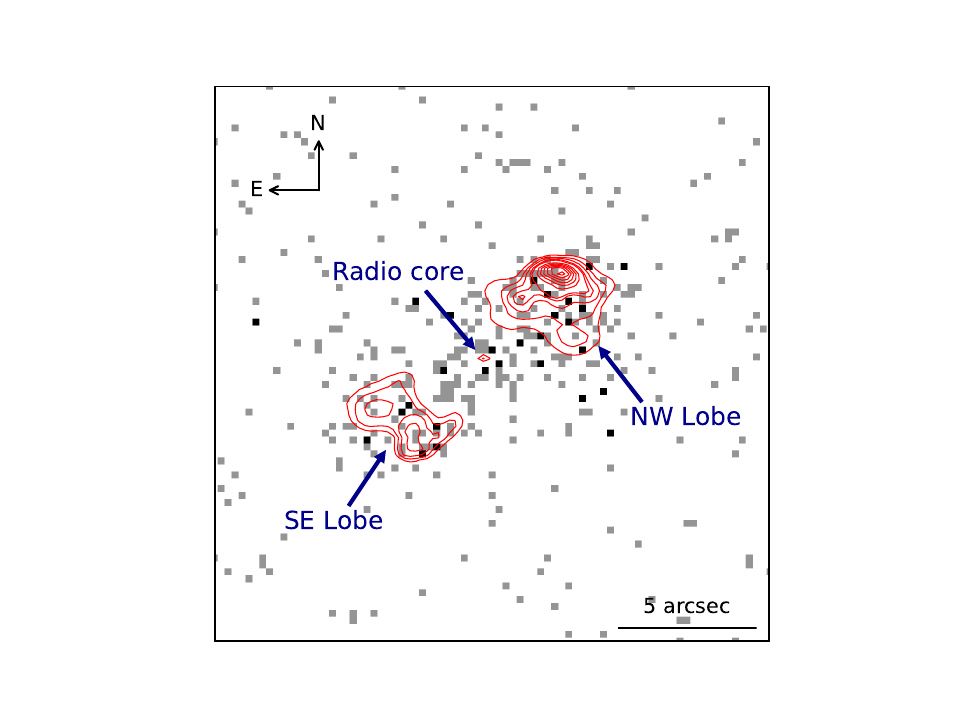}
\caption{\chandra\ 0.5--8~keV 198~ks image of the 3C~220.3 system.  Gray and black squares show X-ray counts at positions from sub-pixel ($0\farcs25 \times 0\farcs25$) binning. Superposed contours show the 9~GHz radio image \citepalias{Haas_et_al} with lobes and core labeled. The image is 20\arcsec\ on a side with north up and east to the left.}
\label{fig:chandra-native_binned}
\end{figure}

In 2014, 3C~220.3 was observed with \chandra~ACIS-S five times for a total exposure time of 198~ks (Table~\ref{table:chandra_data}). The observations were processed, reprojected, and merged with \ciao\ v4.7 and \textsc{caldb} v4.6.4. Like the 2013 \citepalias[their Fig.~5]{Haas_et_al} data, the 2014 \chandra\ image (Figure~\ref{fig:chandra-native_binned}) shows diffuse X-ray emission near and between the radio lobes. Count extraction over the full system region using \texttt{dmextract} in \ciao\ v.4.11 gave $\sim$200 counts. Including the initial, low-count 2013 observation would provide little additional information, and the rest of this paper uses only the deeper 2014 data.

The \chandra\ pointing accuracy is 0\farcs5.\footnote{See Table 5.1 in the \chandra\ Cycle 15 Proposers' Observatory Guide (\url{https://cxc.harvard.edu/proposer/POG/arch_pdfs/POG_cyc15.pdf}).} Lack of sources in the field with both X-ray and visible emission and lack of a clear detection of the X-ray core prevent more accurate alignment of the \chandra\ data.\par

Given that centers of galaxies~A and~B are only ${\sim}$1\farcs5 apart, contamination is a concern. With an average off-axis angle of approximately 0\farcm32, ${\sim}15\%$ of the counts from a point source will lie outside a 1\arcsec\ radius.\footnote{Based on Figure 4.23 in the \chandra \ Proposers' Observatory Guide (\url{http://cxc.harvard.edu/proposer/POG/html/chap4.html}).} For X-ray measurements, we used 1\arcsec-radius circular regions standard to \chandra\ analyses (Regions 2 and 4 in Table~\ref{table:regions_table}), but for better comparison with the \hst\ and Keck imaging, we also measured in 0\farcs75-radius circles. The counts within those regions (Regions~1 and~3, Table~\ref{table:regions_table}) for galaxies A and~B are just below 15 and 7 counts, respectively (Table~\ref{table:Chandra_counts_fluxes_HR_lum}), which correspond to a maximum of 2 to 3 counts outside those regions, assuming unresolved X-ray emission from A and B. The overlap between the 0\farcs75-radius circles of the core and galaxy~B is approximately 0.003~square arcseconds. This results in ${\sim}0.1$ count of contamination between the two regions, a negligible amount.\par

\begin{table}[h]
\caption{2014 \chandra \ exposures}
\label{table:chandra_data}
\centering
\begin{tabular}{lcc}
\hline \hline
OBSID & Date & Exposure (ks) \\
\hline
16081 & 2014-07-03 & 44.49 \\
16082 & 2014-09-15 & 19.81 \\
16520 & 2014-06-16 & 44.71 \\
16521 & 2014-06-18 & 44.49 \\
16522 & 2014-06-30 & 44.49 \\
\hline
\end{tabular}
\tablecomments{Proposal \#15700379, PI Wilkes}
\end{table}

\subsection{MMT Binospec}
\label{subsec:mmt-bino-data}
Optical spectra for galaxy~A, galaxy~B, and the SMG were taken on UT 2021 February 7 with the Binospec instrument \citep{2019PASP..131g5004F} on the 6.5~m MMT telescope at Fred Lawrence Whipple Observatory (FLWO; Mount Hopkins, Arizona). The 20\arcsec$\times$1\arcsec\ slit was oriented along the galaxy~A--galaxy~B axis ($\rm PA = 34\arcdeg$, similar to the Keck spectrum used by \citetalias{Haas_et_al}). Four 20-minute exposures were taken with the 270~lpm grating and central wavelength 6360~\AA\ with average seeing 0\farcs9 (guider camera and wavefront sensor averaged). The spectra covered 3900--9240~\AA\ with spectral resolution $R\approx1340$. A, B, and the SMG have distinguishable traces in the combined spectrum, which is analyzed in Section~\ref{subsec:optical-spectra}. The spectra were processed and reduced with the CfA Binospec pipeline v0.9.9-20180809-rc. 

\subsection{Other Observations}
Table~\ref{table:radio_mm_ir_data} summarizes the radio, submillimeter, and infrared observations used in this paper. Pre-existing data were described by \citetalias{Haas_et_al} (their Section~2).
Assuming a radio spectral index $\alpha = 0.3$ for the core \citep{1997MNRAS.284..541M_core_index} and $\alpha = 0.9$ \citep{1980MNRAS.190..903L_lobe_index} for the lobes, we converted their 9~GHz radio-core flux density of 0.8~mJy to an equivalent 5~GHz flux density to determine a radio core-dominance parameter $\RCD \approx 0.0034$. This \RCD\  translates to an inclination angle of the radio axis to the line-of-sight of $71\degr \pm 10\degr$ \citep[Section 3 of][]{RCD_calcs}.

\begin{table*}[hbt!]
\caption{Radio, submillimeter, and infrared Observations}
\label{table:radio_mm_ir_data}
\centering
\begin{tabular}{lllhll}
\hline \hline
Instrument & Configuration & Frequency or & Bandwidth & Date & Exposure \\
& or Detector &Wavelength&&&time\\
\hline
VLA & A & 9.00~GHz & 2048~MHz & 2012-11-30 & 37~min  \\
SMA & Extended & 302.9~GHz & 8~GHz & 2013-01-11 & 9.6~hr \\
SMA & Compact & 302.9~GHz & 8~GHz & 2013-02-10 & 6.7~hr\\
Keck II & NIRC2 & 2.124~\micron &0.351~\micron & 2012-12-24 & 960~s\\
\hline
\end{tabular}
\tablecomments{All data from \citetalias{Haas_et_al}.}
\end{table*}

\begin{figure}[hbt!]
\centering
\includegraphics[width=0.45\textwidth]{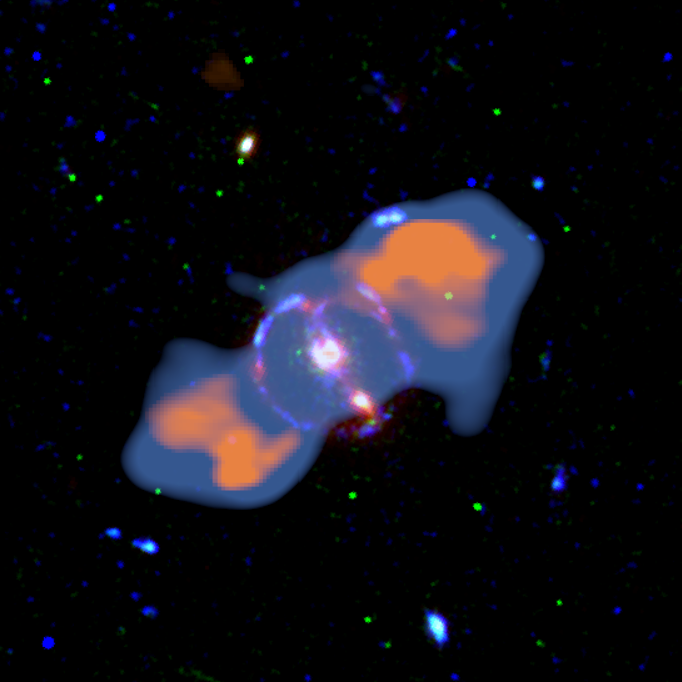}
\caption{Multiwavelength image of the 3C~220.3 system. F606W, F814W, and F160W are shown in blue, green, and red, respectively. \chandra\ data are shown in pale blue-grey, and VLA data are colored orange. An interactive version of this figure is available online. Clicking on the different buttons will switch between visible, X-ray, and radio views. The FITS files provided for the interactive figure include the cutouts of 3C~220.3 in the three \hst\ filters, the 9~GHz VLA radio data, the smoothed cutout of the \chandra\ data, and the full, unsmoothed \chandra\ data. The FITS files can be used for photometry and other morphological analyses.}
\label{fig:composite-image}
\end{figure}

\section{3C~220.3 System Components}
\label{sec:analysis-results}

Combining all multi-wavelength components (i.e., galaxies A and B, radio lobes, Einstein ring), the 3C~220.3 system has approximate dimensions of $11\arcsec \times 6\arcsec$ with  position angle (PA) of the long (radio lobe) axis at 130\degr\ and  PA of the shorter axis (intersecting the Einstein ring, galaxy~A, and galaxy~B) at 40\degr. The northwest and southeast radio lobes have dimensions $4\farcs5 \times 3\farcs5$ (32~kpc $\times$ 25~kpc) and $4\arcsec \times 3\arcsec$ (29~kpc $\times$ 22~kpc), respectively, with the centers of both lobes lying ${\sim}3\farcs5$ (25~kpc) from the radio core. Figure~\ref{fig:composite-image} shows a color-coded, multi-wavelength overview of the system.

\begin{figure*}[hbt!]
\includegraphics[width=\textwidth,trim=3cm 2.16cm 2.4cm 2.3cm, clip]{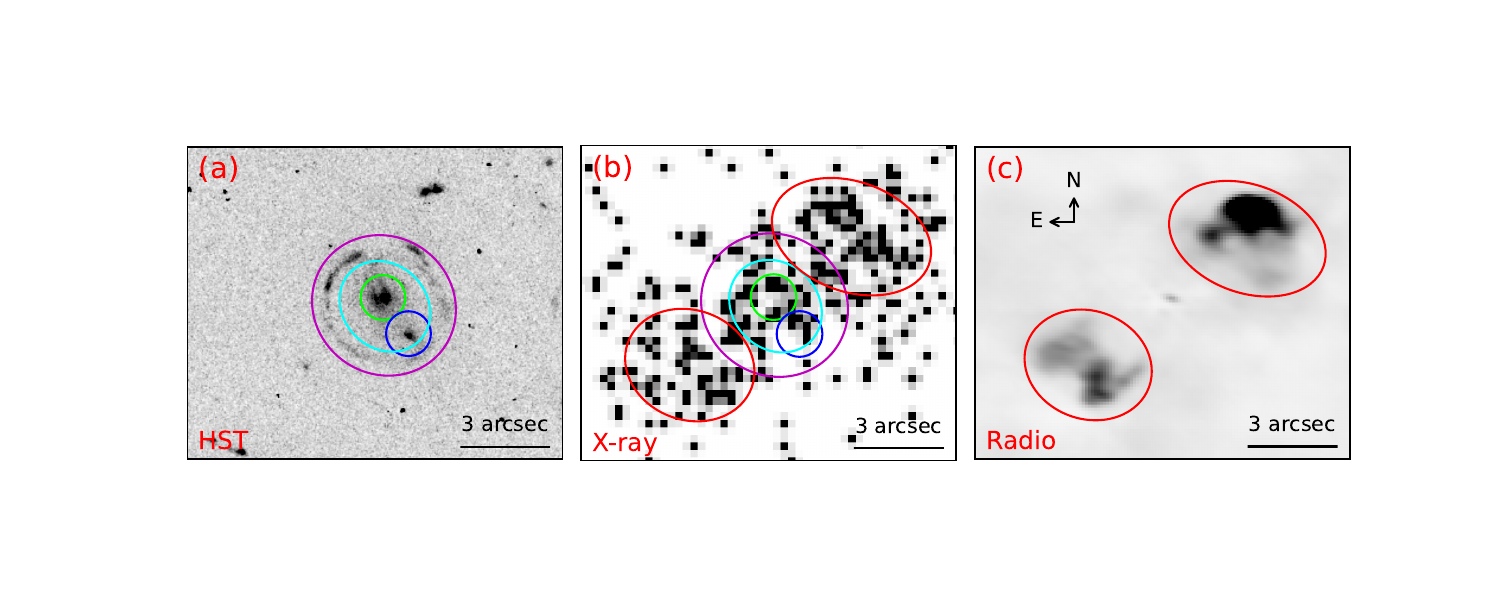}
\caption{Negative images of the 3C~220.3 system. Left to right are visible (\hst\ F606W+F814W), \chandra\ X-ray (0.5--8~keV,
$0\farcs25 \times 0\farcs25$ sub-pixel binned, Gaussian smoothed with $r=1$~pix),
and VLA 9.0~GHz. Images are $12\farcs6 \times 10\farcs5$. Extraction regions shown, defined in Table~\ref{table:regions_table},  are 1 (galaxy~A, green circle), 3 (galaxy~B, blue circle), 5 (extended galaxy, cyan ellipse),  6 (Einstein ring, annulus between magenta and cyan ellipses), and 7 and 8 (NW and SE radio lobes, red ellipses). Region 9 (Table~\ref{table:regions_table}, not show in this figure) measures the X-ray emission from the full source to allow comparison with  lower-resolution X-ray data on this source.}
\label{fig:extraction_regions}
\end{figure*}

\begin{deluxetable*}{clllll}
\caption{Region parameters for flux analyses}
\label{table:regions_table}
\tablehead{\colhead{No.} & \colhead{Region} & \colhead{Region type} & \colhead{Position (R.A. and Decl., FK5)} 
& \colhead{Radii\tablenotemark{a}} & \colhead{PA\tablenotemark{b}}}

\startdata
1 & Galaxy~A (small)  & Circle & 9:39:23.8509 +83:15:25.864 & 0\farcs75 & \\ 
2 & Galaxy~A (large) & Circle & 9:39:23.8509 +83:15:25.864 & 1\farcs00 &  \\ 
3 & Galaxy~B (small) & Circle & 9:39:23.3666 +83:15:24.653 & 0\farcs75 & \\ 
4 & Galaxy~B (large) & Circle & 9:39:23.3666 +83:15:24.653 & 1\farcs00 & \\ 
5 & Ext gal\tablenotemark{c} (no A or B) & \makecell[l]{Ellipse with 2\\circle exclusions} & \makecell[l]{9:39:23.8134 +83:15:25.567\\{[9:39:23.8509 +83:15:25.864]}\\{[9:39:23.3666 +83:15:24.653]}} & \makecell[l]{1\farcs63/1\farcs40\\{[0\farcs75]}\\{[0\farcs75]}} & \makecell[l]{314\degr\\~\\~} \\[4.0ex]
5a & Ext gal\tablenotemark{c} + A (no B) & \makecell[l]{Ellipse with 1\\circle exclusion} & \makecell[l]{9:39:23.8134 +83:15:25.567\\{[9:39:23.3666 +83:15:24.653]}} & \makecell[l]{1\farcs63/1\farcs40\\{[0\farcs75]}} & \makecell[l]{45\degr\\~} \\[3.0ex] 
5b & Ext gal\tablenotemark{c} (with A \& B) & Ellipse & 9:39:23.8134 +83:15:25.567 & 1\farcs63/1\farcs40 & 45\degr \\[2.0ex]
6 & Einstein ring\tablenotemark{c} & \makecell[l]{Elliptical\\annulus} & \makecell[l]{9:39:23.8321 +83:15:25.600\\ {[9:39:23.8134 +83:15:25.567]}} & \makecell[l]{2\farcs46/2\farcs30\\ {[1\farcs63/1\farcs40]}} & \makecell[l]{235\degr \\ {[45\degr]}} \\[2.0ex]
7 & NW lobe & Ellipse & 9:39:22.3978 +83:15:27.843 & 2\farcs70/1\farcs80 & 70\degr \\
8 & SE lobe & Ellipse & 9:39:25.4113 +83:15:23.636 & 2\farcs16/1\farcs80 & 70\degr \\
9 & Full X-ray source\tablenotemark{d} & Box & 9:39:23.8509 +83:15:25.86 & 11\farcs25/5\farcs79 & 124\degr \\[0.4ex]
\tableline
 & \makecell[l]{\\Background\\~~~(\chandra)\tablenotemark{c}} & \makecell[l]{\\Elliptical\\annulus} & \makecell[l]{\\9:39:23.6892 +83:15:25.621\\ {[9:39:23.6892 +83:15:25.621]}} & \makecell[l]{\\18\farcs90/9\farcs00\\ {[9\farcs00/5\farcs40]}} & 130\degr \\[4.0ex]
 & \makecell[l]{Background\\~~~(\hst, Keck)} & Circles & \makecell[l]{9:39:26.5319 +83:15:32.506\\ 9:39:24.3255 +83:15:34.181\\ 9:39:29.1934 +83:15:26.008\\ 9:39:24.2492 +83:15:18.240\\ 9:39:20.4475 +83:15:16.699\\ 9:39:20.0645 +83:15:34.683\\ 9:39:19.3807 +83:15:29.458\\ 9:39:18.9060 +83:15:25.841\\ 9:39:28.8691 +83:15:21.052} & 0\farcs67 & \\\\
 & \makecell[l]{Background\\~~~(\hst \ F702W)} & Circles & \makecell[l]{9:39:26.7921 +83:15:30.756\\ 9:39:24.3255 +83:15:34.181\\ 9:39:27.8687 +83:15:27.092\\ 9:39:24.2492 +83:15:18.240\\ 9:39:20.9199 +83:15:21.699\\ 9:39:21.8634 +83:15:32.100\\ 9:39:19.8540 +83:15:28.958\\ 9:39:18.9060 +83:15:25.841\\ 9:39:27.0244 +83:15:22.803} & 0\farcs67 & \\
\enddata
\tablenotetext{a}{Pairs of numbers separated by a slash indicate the semi-major and semi-minor axes of an ellipse or the length and width of a box.}
\tablenotetext{b}{Position Angle, degrees east of north}
\tablenotetext{c}{Ext gal refers to extended emission around galaxy~A as shown by the cyan circle in Figure~\ref{fig:extraction_regions}. Einstein ring refers to the annulus between the cyan and magenta circles in Figure~\ref{fig:extraction_regions}. For annular regions, the first line refers to the outer shape, and additional lines in brackets refer to excluded inner shapes.}
\tablenotetext{d}{Region 9 measures the X-ray emission from the full source to allow direct comparison with  lower-resolution X-ray data.}
\tablecomments{DS9 region files are available from corresponding author upon request.}
\end{deluxetable*}

\subsection{Observed Visible and Infrared Photometry Analysis}
\label{subsec:oir-analysis}

The \hst\ images (Figure~\ref{fig:new-hst-einstein-rings}) show three main components: galaxy~A, galaxy~B, and the SMG. There is also extended emission around galaxy~A, most easily seen in Figure~\ref{fig:new-hst-einstein-rings}b. (Rows 1--6 of Table~\ref{table:regions_table} give positions of these components.)  The Einstein ring from the lensed SMG is visible in all images, including the NIR image.  The ring's radius is 2\farcs0, overlapping the inner parts of the radio lobes. The center of the ring is ${\sim}0\farcs2$ to the southwest of the centroid of galaxy~A. Galaxy~B is ${\sim}1\farcs5$ from galaxy~A and is not an image of the lensed SMG, as it lies (${\sim}0\farcs7$) inside the edge of the Einstein ring.

For the \hst\ and Keck images, counts (in units of electrons per second) were extracted using the Funtool \texttt{funcnts} (``Counts in Regions'') in \ciao\ v4.11 of \dsn.
Figure~\ref{fig:extraction_regions} plots the regions for observed optical, IR, and X-ray count extractions, and Table~\ref{table:regions_table} defines them, along with the regions used to derive the sky background in each image.  The backgrounds were subtracted from the counts in regions, and the resulting net counts were converted to flux densities using the filter-specific inverse sensitivity factors (\texttt{PHOTFNU} or \texttt{PHOTFLAM}) listed in the FITS headers for each filter.
The counts and flux densities are listed in Table~\ref{table:HST_Keck_counts_fluxes}. \par

To maintain consistency, we redid the photometry for the Keck image using our regions (Table~\ref{table:regions_table}), which differ slightly from those reported by \citetalias{Haas_et_al}. We converted to flux density (Table~\ref{table:HST_Keck_counts_fluxes}) using the same calibration source (2MASS 09393423+8314518, $K_s = 15.04 \pm 0.124$) as \citetalias{Haas_et_al} (their Sec.~2.4). 

\subsection{Observed Optical Spectroscopy Analysis}
\label{subsec:optical-spectra}

\label{subsubsec:spatial-prof}
\begin{figure*}[hbt!]
\centering
\includegraphics[width=\textwidth]{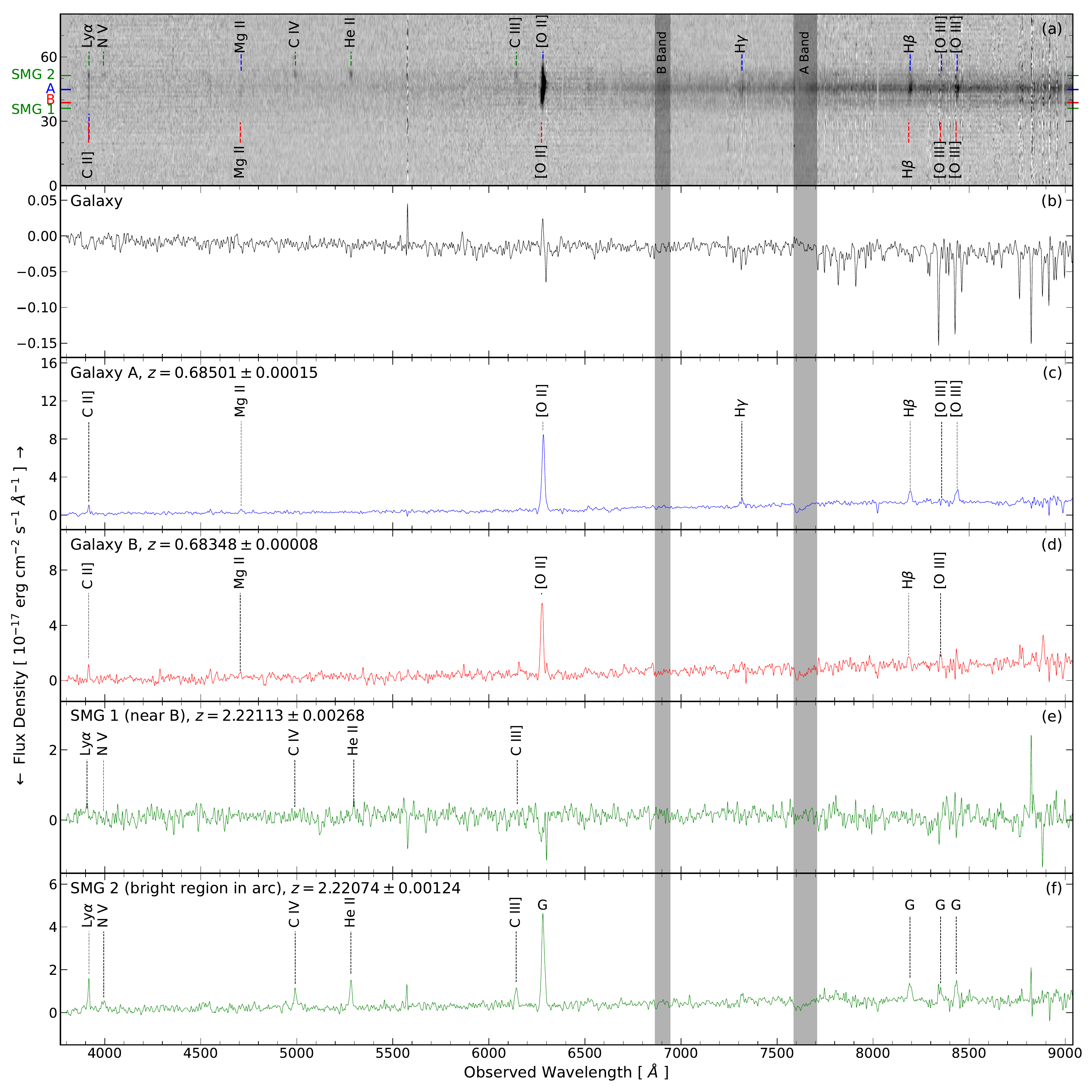}
\caption{Top: two-dimensional spectrum from MMT Binospec. Observed wavelength increases to the right. The vertical axis corresponds to the sky direction of the slit (Figure~\ref{fig:new-hst-einstein-rings}a), with the bottom corresponding to the southwest end of the slit.
Component positions are marked at the left edge. Bottom five panels: extracted spectra of (top to bottom) the extended galaxy, galaxy~A, galaxy~B, the SW part of the SMG near B, and the NE part of the SMG. Emission lines are marked in each panel. Lines in panel (f) marked by a {\sffamily G} are likely contamination from the extended galaxy or galaxy~A.
There is strong \Oii\ emission for both galaxy~A and galaxy~B. \Oiii, \Hbeta, \Mgii, and \Cii\ emission are also detected in both A and B spectra.}
\label{fig:bino-spectra}
\end{figure*}

\begin{figure}[hbt!]
\includegraphics[width=0.46\textwidth]{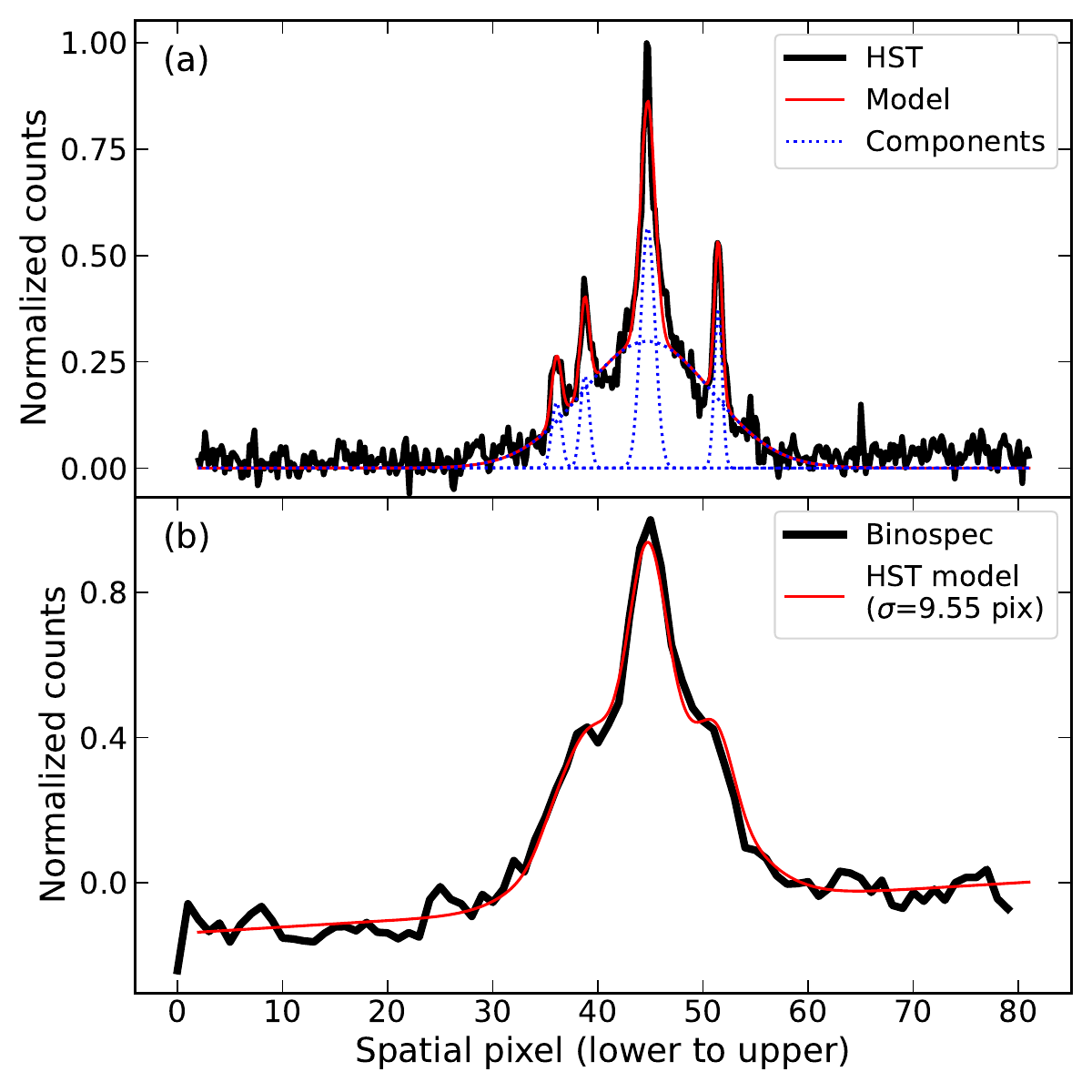}
\caption{Top: Expected flux distribution along the Binospec slit based on the \hst\ F606W+F814W image (Figure~\ref{fig:new-hst-einstein-rings}a). The solid black line shows the \hst\ profile, the dotted blue lines show the five Gaussian model components, and the solid red line shows their sum.  The strongest component is galaxy~A, with galaxy~B to its left (southwest).  The two outer peaks are the Einstein ring of the SMG, and the fifth component is a broad Gaussian corresponding to the extended emission. Bottom: Comparison of the blurred \hst\ spatial model (from above, red) to the observed MMT Binospec spatial profile summed over all wavelengths (black).}
\label{fig:bino-hst-analysis}
\end{figure}

The two-dimensional 2021 MMT Binospec spectrum (Figure~\ref{fig:bino-spectra}, top) shows four clear traces that correspond to galaxies~A and~B and the two spots where the Einstein ring of the SMG crossed the slit (Figure~\ref{fig:new-hst-einstein-rings}a).
Given these sources' proximity, however, contamination between them precludes simply summing over a range of pixels to extract spectra for individual galaxies or components. We therefore matched the Binospec slit to the 2014 \hst\ image (as in Figure \ref{fig:new-hst-einstein-rings}a) to produce a sub-arcsecond ``sky image" of the slit. The \hst\ spatial profile produced from summing along the length of this slit cutout reveals four sharp peaks (Figure~\ref{fig:bino-hst-analysis}a) corresponding to the spectral traces. As Figure~\ref{fig:bino-hst-analysis}a shows, there is also an extended component (``Galaxy'' in Figure~\ref{fig:bino-spectra}), which we modeled as a Gaussian. All five components were fit  using the \textsc{lmfit} package \citep{2021zndo...4516651N}. Convolving the best-fit model with a Gaussian kernel based on the average seeing during the Binospec observations provides a good match between the \hst\ model and the Binospec spatial profile (Figure~\ref{fig:bino-hst-analysis}b).

\label{subsubsec:spec-extract}
If the spatial profile were constant over the Binospec wavelength range, the blurred spatial profile model would give a measure of the contribution of each of the five components per spatial pixel such that the value of pixel $i$ is given by 
\begin{equation}
    p_i = p_{i,\text{gal}} +  p_{i,\text{A}} +  p_{i,\text{B}} +  p_{i,\text{SMG1}} +  p_{i,\text{SMG2}}\;.
\end{equation}
Based on these per-pixel contributions, we formed an $80 \times 5$ weights matrix $w$ (80 spatial Binospec pixels and 5 components), such that the dot product of the weights matrix and the true spectra of all five components ($5 \times 4087$ matrix $f$) produces the two-dimensional (2D) spectrum observed by Binospec ($80 \times 4087$ matrix $D$). In other words,
\begin{equation}
    w \cdot f = D\;.
\end{equation}
With the matrices $D$ and $w$ known, we can multiply by the inverse of the weights matrix ($w^{-1}$) to solve for the true-spectra matrix ($f$):
\begin{equation}
    w^{-1} w ~ f = w^{-1} D \Rightarrow f = w^{-1} D \;.
\label{eq:inv}
\end{equation}
As the weights matrix is non-invertible ($80 \times 5$ rather than square), $w^{-1}$ is actually the left inverse (specifically, a Moore--Penrose pseudoinverse). Equation~(\ref{eq:inv}) gives the five spectra shown in Figure~\ref{fig:bino-spectra}b--f.

\paragraph{\textbf{Galaxy~A}}
\citetalias{Haas_et_al}'s optical spectroscopy yielded detections of \Cii\,$\lambda$2326, \Oii\,$\lambda$3727, \Hbeta, the \Oiii\ doublet, and weak \Mgii. Our extracted galaxy~A spectrum from the Binospec data shows those same lines, as well as H$\gamma$. The redshift \zAErr\ determined from emission line fitting is consistent with previous values (\citetalias{Haas_et_al}, \citealt{1985PASP...97..932S}). 
The H$\beta$ line is broader than the instrumental resolution with $\rm FHWM=340\pm50$~\kms.  Other lines have signal-to-noise ratios (S/N) that are too low to determine widths, which could be unresolved or as broad as H$\beta$ within the uncertainties.
The \Oii\ line is ${\sim}3$~times stronger than the \Oiii\ and \Hbeta\ lines. This is somewhat unusual for radio galaxies, which generally have \Oii/\Oiii\,$\lambda5008$ line ratios around 0.2 \citep{1990agn..conf...57Netzer,2000asqu.book..585W}. Strong \Oii\ emission is often used as an indicator for star formation and starburst galaxies.

\paragraph{\textbf{Galaxy~B}}
\citetalias{Haas_et_al} noted a possible blue-shift of ${\sim}5$~\AA\ relative to galaxy~A for the \Oii\ and \Hbeta\ lines in galaxy~B's part of the trace, but with their signal-to-noise and seeing ($\sim$1\farcs5), they could not be certain. The excellent seeing of our 2021 Binospec observations gave a clear spectral trace of galaxy~B, distinct from galaxy~A. In addition to the \Oii\ emission line, we also detected \Cii, \Hbeta, and the \Oiii\ doublet  (Figure~\ref{fig:bino-spectra}c), which suggest possible AGN activity. Fitting the emission lines gave \zBErr, corresponding to a (blue-shifted) wavelength difference of 5.7~\AA. The S/N was too low to determine any line widths.

\begin{figure*}[hbt!]
\centering
\includegraphics[width=\textwidth]{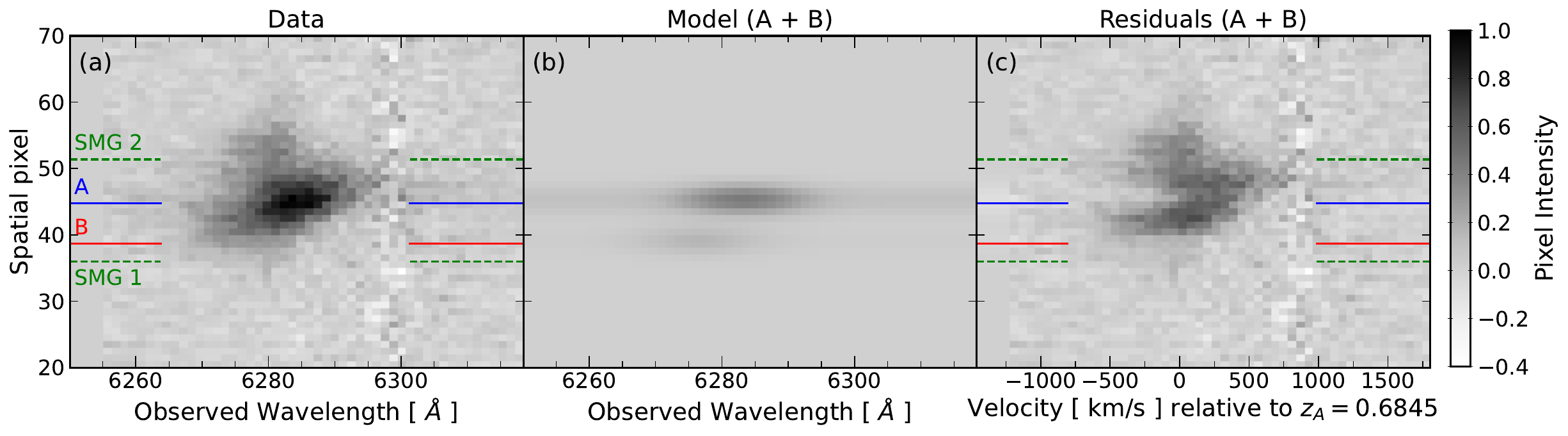}
\caption{Close-up of the \Oii\ emission line in the two-dimensional spectrum.  Horizontal axis is wavelength for panels (a) and (b) and velocity relative to the redshift of galaxy A (\zA) in panel (c). The vertical axis is spatial position along the slit from SW to NE (Figure~\ref{fig:new-hst-einstein-rings}a). The  scale, in 0\farcs24 pixels, is shown at the left. Panels left to right show the data, the galaxy~A and~B components of the 2D model, and the residuals after subtracting the galaxy~A and~B components from the data. Galaxy positions along the slit are marked in panels (a) and (c).}
\label{fig:OII-structure}
\end{figure*}

\paragraph{\textbf{Extended galaxy}}
The spectrum for the extended galaxy (Figure~\ref{fig:bino-spectra}b) is virtually featureless, showing only telluric contamination and what seems to be ``leakage'' from the other components. 
\label{subsubsec:oii-structure}
However, a closer look at the \Oii\ line reveals a complicated spatial structure (Figure~\ref{fig:OII-structure}a). To better understand the underlying structure, we fit the 2D spectrum around the \Oii\ line with a two-component model comprising 2D Gaussians for galaxies A and B. We fixed the components' positions and widths in both the spatial and wavelength dimensions (Section~\ref{subsubsec:spatial-prof}), allowing only their amplitudes to vary. Even with an apparent over-subtraction of galaxy~A (Figure~\ref{fig:OII-structure}), there is clear evidence of an extended component of \Oii\ emission. In particular, there is additional light at negative relative velocity extending along the spatial pixel axis, and there is a bright spot at positive velocity that neither galaxy~A nor galaxy~B can account for. The extended component looks irregular (Figure~\ref{fig:OII-structure}c), and  a Gaussian would be a poor fit to its shape.

\label{subsubsec:z-or-dop}
The redshift of galaxy~B is lower than that of galaxy~A by 459~\kms\ in the rest frame (if the galaxies are at the same distance). The projected  separation is 10.7~kpc. Combined, these values suggest a merging or interacting system \citep{2009A&A...498..379D,2011ApJ...742..103L}. 
An interaction would likely trigger AGN activity and star formation. Star formation in a merging system could in turn explain the unusual structure of the \Oii\ line (Figure~\ref{fig:OII-structure}), the unusually high \Oii\ emission from galaxy~A, and the diffuse optical emission that makes up both the extended galaxy and the tail-like structure to the northwest of galaxy~A seen in the \hst\ images (Figure~\ref{fig:new-hst-einstein-rings}).

\paragraph{\textbf{SMG}}
The ``SMG1'' spectrum (Figure~\ref{fig:bino-spectra}e) corresponds to the part of the arc nearest galaxy~B. This emission is very faint and results in a virtually featureless spectrum in terms of emission lines. The ``SMG2'' spectrum (Figure~\ref{fig:bino-spectra}f) corresponds to the bright arc in the northeast part of the Einstein ring and is the section that \citetalias{Haas_et_al} detected in their Keck spectrum. In addition to the SMG lines reported by \citetalias{Haas_et_al} (Ly$\alpha$, \ion{C}{4}\,$\lambda$1549, \ion{He}{2}\,$\lambda$1640), we detected \ion{N}{5}\,$\lambda$1241 and \ion{C}{3}]\,$\lambda$1909. Fitting these emission lines gives \zSMGBErr\ for the SMG, consistent with the value reported by \citetalias{Haas_et_al}. The SMG2 spectrum also shows \Oii, \Hbeta, and \Oiii\ emission lines at $z=0.68527 \pm 0.00347$, consistent with the redshift of galaxy~A. These lines are presumably ``leakage'' from galaxy~A or perhaps the extended galaxy. As mentioned in at the start of this subsection, the spectrum extractions assume that the spatial profile is constant over all wavelengths, but this is not a correct assumption, as seen in the spectral traces in Figure~\ref{fig:bino-spectra}a. Higher-resolution, \Oii-specific imaging  is necessary to extract accurate spectra for the individual components.


\subsection{X-Ray Analysis}
\label{subsec:xray-analysis}

The \chandra\ data re-sampled to 0\farcs25 bins are shown in   Figure~\ref{fig:extraction_regions} with the primary regions superposed. There is no X-ray peak at the radio core, and only $19.6 \pm 4.6$ net counts lie within a 1\arcsec~extraction region (Galaxy~A, large, Table~\ref{table:Chandra_counts_fluxes_HR_lum}). The low counts and low ratio of X-ray to total radio emission of ${\sim}{-}1.6$~dex are  consistent with the AGN being highly obscured or Compton thick. The ratio of X-ray to total radio emission for Compton-thick sources is $\log L_{X, 0.5-8~\text{keV}}/L_{178~\text{MHz}}<0$ \citep{Kuras2021}.

\label{subsec:behr-hardness}

For low-count X-ray sources like 3C~220.3,  hardness ratios ($\mathcal{HR}$) are useful to characterize the X-ray emission. We adopted standard definitions $S$ = counts in range 0.5--2~keV, and $H$ = counts in range 2--8~keV for the soft and hard bands, respectively, and used the program \texttt{BEHR} (Bayesian Estimation of Hardness Ratios, \citealt{BEHR_paper}) to calculate hardness ratios ($\mathcal{HR}\equiv(H-S)/(H+S)$) for the extraction regions given in Table~\ref{table:regions_table}. 
Results for all our defined regions are in Table~\ref{table:Chandra_counts_fluxes_HR_lum}. All of the entries are negative, showing soft SEDs overall. For comparison, an unobscured AGN with a typical slope of $\Gamma \sim 1.9$ (i.e., a spectral index of $\alpha=0.9$ with $\Gamma=1+\alpha$) will have $\mathcal{HR} \sim -0.5$. However, NLRGs cover a wide range of hardness ratios: $-0.7 < \mathcal{HR} < 0.7$ \citep{2013_Wilkes_obscured_3CRR}. Although the nuclei of edge-on NRLGs are highly obscured, the X-ray emission often appears softer than expected \citep{2013_Wilkes_obscured_3CRR,Kuras2021}, as it does in 3C~220.3.
The softer X-ray emission comes from additional components near the AGN core such as scattered light or emission from an extended, unobscured region. These would be minor contributors to an unobscured AGN, but they become important when the AGN itself is obscured.

\label{subsec:xray-spectral-analysis}
Since 3C~220.3 has so few X-ray counts, spectral models for individual regions such as the core or galaxy~B are uncertain. Radio galaxies have a typical X-ray photon index $\Gamma = 1.9$ \citep{agn_photon_index}, which is consistent with 3C~220.3's radio spectral index (Section~\ref{subsec:sed}) and with the X-rays being generated by Compton scattering. Fixing $\Gamma = 1.9$ allows the \ciao\ program \texttt{Sherpa} \citep{sherpa:freeman2001a,sherpa:freeman2001b,sherpa:doe2007,sherpa:burke2020} to  estimate the 0.5~keV to 8~keV  X-ray fluxes. For these fits, we assumed a standard power law model, $N{\left(E)\propto A(E/{E}_{{\rm{ref}}}\right)}^{-{\rm\Gamma }}\,\exp [-{N}_{{\rm{H}}}\,\sigma (E)]$, with 3C~220.3's Galactic absorption  $N_{\rm H} = 3.26 \times 10^{20}$~{atoms~cm}$^{-2}$.
Given the low signal, particularly at the ends of the 0.5--8~keV energy window, we tested spectral fits with different energy binning. The results are in good agreement and were averaged to determine the best-fit fluxes for each region. Table~\ref{table:Chandra_counts_fluxes_HR_lum} shows the results.

From the observed counts in the region of the AGN core (Galaxy~A), the derived X-ray luminosity of 3C~220.3 is at the lower end of the typical AGN range of $10^{42}$--$10^{46}$~erg~s\textsuperscript{-1} \citep{AGN_luminosity}. 3C~220.3 (Table~\ref{table:Chandra_counts_fluxes_HR_lum}) is viewed edge-on and known to be heavily obscured, thus it is likely that the intrinsic X-ray luminosity of the AGN core is around 1000~times higher \citep{Kuras2021}. Since the observed hardness ratio is soft, showing no indication of the core's obscuration, it is more likely that the X-ray emission in this region is dominated by the diffuse extended component and that the AGN contribution is weak or undetected. 

The X-ray luminosity of ${\sim}10^{42}$~erg~s\textsuperscript{-1} in the region centered on galaxy~B is high for anything other than an AGN. If the counts originate in Galaxy~B, the combination with a potential \Cii\ detection in the Binospec spectra (Section~\ref{subsec:optical-spectra}) suggests that the galaxy may host a low luminosity or heavily obscured AGN. There is no radio detection of galaxy~B. Alternatively, the X-ray counts from this region may be dominated by the diffuse emission in the region, similarly to Galaxy A. With the current data, it is not possible to distinguish these two scenarios (but see Section~\ref{subsec:BAYMAX}).
Thus, Table~\ref{table:regions_table} region 5b (including the circles for Galaxies A, B) is the most relevant region for estimating the X-ray emission in between the radio lobes. 

\movetabledown=1cm
\begin{deluxetable*}{clCCCCCCCCCC}
\rotate
\tablecaption{Extracted counts, hardness ratios, and averaged Sherpa fluxes/luminosities from \chandra \ X-ray data\label{table:Chandra_counts_fluxes_HR_lum}}

\tablehead{ & & \colhead{Counts\tablenotemark{b}} & \colhead{Hardness} & \colhead{X-ray flux\tablenotemark{d,e,f}} & \colhead{X-ray luminosity\tablenotemark{d,g}} & \colhead{1~keV flux density}\\ 
\colhead{No.} & \colhead{Region\tablenotemark{a}} & \colhead{(photons)} & \colhead{Ratio\tablenotemark{c}} & \colhead{$10^{-15}$~erg~cm$^{-2}$~s$^{-1}$} & \colhead{$10^{42}$~erg~s$^{-1}$} & \colhead{$10^{-16}$~erg~cm$^{-2}$~s$^{-1}$~keV$^{-1}$}}

\startdata
1 & Galaxy~A\tablenotemark{\small h} (small) & 14.2 \pm 3.9 & -0.6_{-0.2}^{+0.2} & 0.7_{-0.5}^{+0.2} & 1.5_{-1.1}^{+0.5} & 2.3 \pm 1.7 \\
2 & Galaxy~A\tablenotemark{\small h}(large) & 19.6 \pm 4.6 & -0.4_{-0.2}^{+0.2} & 1.2_{-0.7}^{+0.6} & 2.5_{-1.4}^{+1.3} & 3.8_{-2.1}^{+2.0} \\
3 & Galaxy~B\tablenotemark{\small h}(small) & 6.2 \pm 2.6 & -0.5_{-0.5}^{+0.2} & 0.3_{-0.3}^{+0.4} & 0.6_{-0.6}^{+0.8} & 0.7_{-0.7}^{+1.1} \\
4 & Galaxy~B\tablenotemark{\small h}(large) & 8.6 \pm 3.2 & -0.5_{-0.5}^{+0.2} & 0.6_{-0.5}^{+0.5} & 1.3_{-1.0}^{+1.0} & 1.9_{-1.4}^{+1.5} \\
5 & Ext gal (no A or B) & 20.1 \pm 4.7 & -0.2_{-0.2}^{+0.2} & 1.9_{-1.3}^{+1.2} & 4.0_{-2.7}^{+2.6} & 6.1_{-4.2}^{+3.9} \\
5a & Ext gal + A (no B) & 32.3 \pm 6.1 & -0.4_{-0.2}^{+0.1} & 1.6_{-0.8}^{+0.8} & 3.4_{-1.6}^{+1.7} & 5.0_{-2.3}^{+2.4} \\
5b & Ext gal (with A \& B) & 40.8 \pm 6.6 & -0.4_{-0.2}^{+0.1} & 1.7_{-0.8}^{+0.8} & 3.7_{-1.7}^{+1.7} & 5.6_{-2.6}^{+2.6} \\
6 & Einstein ring & 32.3 \pm 6.1 & -0.7_{-0.2}^{+0.2} & 1.5_{-1.2}^{+1.2} & 59.7_{-48.4}^{+46.1} & 4.8_{-3.9}^{+3.7} \\
7 & NW lobe & 65.3 \pm 8.5 & -0.3_{-0.1}^{+0.1} & 2.1_{-1.0}^{+1.1} & 4.3_{-2.0}^{+2.2} & 6.5_{-3.1}^{+3.4} \\
8 & SE lobe & 44.6 \pm 7.1 & -0.4_{-0.2}^{+0.1} & 1.7_{-0.8}^{+0.8} & 3.6_{-1.6}^{+1.7} & 5.5_{-2.4}^{+2.6} \\
9 & Full source & 212.1 \pm 15.3 & -0.4_{-0.1}^{+0.1} & 10.2_{-1.9}^{+1.9} & 21.4 \pm 3.9 & 37.3 \pm 3.8 \\
\enddata
\tablenotetext{a}{See Table~\ref{table:regions_table} for region parameters.}
\tablenotetext{b}{X-ray background count rate: $0.11 \pm 0.01$ photons/pixel; pixel scale: 1~px = 0\farcs5.}
\tablenotetext{c}{Median value and asymmetric uncertainties.}
\tablenotetext{d}{X-ray fluxes and luminosities averaged over results from broadband Sherpa analyses (0.5--8~keV) for binnings of 3, 5, and 7.}
\tablenotetext{e}{Mean value and 68\% confidence interval error bars}
\tablenotetext{f}{Fluxes calculated assuming a power-law spectral form  with $\Gamma = 1.9$ (which is typical of X-ray emission from AGN) corrected for the Galactic $N_H$ towards 3C~220.3 of $3.26 \times 10^{20}$ cm$^{-2}$ (calculated using the Colden Galactic Neutral Hydrogen Density Calculator with NRAO data compilation by \citet{Dickey_Lockman_1990_NH}; \url{http://cxc.harvard.edu/toolkit/colden.jsp}).}
\tablenotetext{g}{Luminosities calculated using luminosity distances $D_{L,z=0.685} = 1.29 \times 10^{28}$ cm and $D_{L,z=2.221} = 5.56 \times 10^{28}$ cm. The luminosity for the SMG is magnified by the gravitational lensing.}
\tablenotetext{h}{The spectral fits for Galaxy~A and Galaxy~B assume an unobscured AGN. However, we conclude  (Section~\ref{subsec:xray-spectral-analysis}) that it is likely the X-ray emission in both regions is dominated by the extended, diffuse component described by region 5b.}
\end{deluxetable*}

\begin{deluxetable*}{clhhhhhCCCCC}
\tablecaption{Visible and near-infrared flux densities for components of the 3C~220.3 system\label{table:HST_Keck_counts_fluxes}}

\tablehead{\colhead{No.\tablenotemark{a}} & \colhead{Region} & \nocolhead{F606W} & \nocolhead{F702W} & \nocolhead{F814W} & \nocolhead{F160W} & \nocolhead{Keck $K'$} & \colhead{F606W} & \colhead{F702W\rlap{\tablenotemark{b}}} & \colhead{F814W} & \colhead{F160W} & \colhead{Keck $K'$} 
}
\startdata
1 & Galaxy~A (0\farcs75) & 16.1 \pm 4.1 & 1.5 \pm 1.4 & 24.3 \pm 5.0 & 231.5 \pm 15.2 & 2869.4 \pm 53.2 & 2.2 \pm 0.06 & 4.6 \pm 0.3 & 8.0 \pm 0.1 & 35.2 \pm 0.1 & 49.3 \pm 1.3 \\
3 & Galaxy~B (0\farcs75) & 7.6 \pm 2.9 & 0.9 \pm 1.2 & 11.8 \pm 3.6 & 124.5 \pm 11.2 & 1622.4 \pm 39.8 & 1.0 \pm 0.06 & 2.8 \pm 0.3 & 3.9 \pm 0.1 & 18.9 \pm 0.1 & 27.9 \pm 1.3 \\
5 & Ext gal (no A or B) & 13.9 \pm 4.0 & 1.1 \pm 1.7 & 15.9 \pm 4.3 & 131.4 \pm 11.5 & 2017.1 \pm 43.7 & 1.9 \pm 0.3 & 3.5 \pm 0.5 & 5.2 \pm 0.4 & 20.0 \pm 0.4 & 34.7 \pm 2.2 \\
5a & Ext gal + A (no B) & 30.1 \pm 5.8 & 2.6 \pm 2.3 & 40.2 \pm 6.6 & 362.9 \pm 19 & 4886.5 \pm 68.7 & 4.1 \pm 0.3 & 8.0 \pm 0.6 & 13.2 \pm 0.5 & 55.1 \pm 0.5 & 84.0 \pm 2.7 \\
5b &Ext gal (with A \& B) & 36.3 \pm 6.4 & 3.3 \pm 2.6 & 50.1 \pm 7.4 & 470.9 \pm 21.7 & 6217.1 \pm 77.4 & 4.9 \pm 0.3 & 10.3 \pm 0.7 & 16.5 \pm 0.5 & 71.5 \pm 0.5 & 106.8 \pm 3.0\0 \\
6 & Einstein ring & 41.4 \pm 7.1 & 2.4 \pm 2.9 & 34.2 \pm 6.5 & 363.0 \pm 19.0 & 4500.4 \pm 64.1 & 5.7 \pm 0.4 & 7.6 \pm 0.8 & 11.3 \pm 0.6 & 55.1 \pm 0.6 & 77.3 \pm 3.6 \\
7\rlap{\tablenotemark{c}} & NW lobe & 22.9 \pm 6.1 & 1.2 \pm 3.4 & 20.0 \pm 5.9 & 158.8 \pm 12.6 & 1974.7 \pm 36.1 & 3.1 \pm 0.5 & 3.8 \pm 1.0 & 6.6 \pm 0.8 & 24.1 \pm 0.7 & 33.9 \pm 4.4 \\
8\rlap{\tablenotemark{c}} & SE lobe & 8.4 \pm 4.4 & 0.3 \pm 2.8 & 4.2 \pm 3.8 & 67.2 \pm 8.2 & 745.5 \pm 16.3 & 1.1 \pm 0.4 & 1.0 \pm 0.9 & 1.4 \pm 0.7 & 10.2 \pm 0.7 & 12.8 \pm 4.0 \\
\tableline
\multicolumn{2}{l}{$10^4\times$background rate\tablenotemark{d}}&&&&&& $6.95 \pm 7.22 $& $7.73 \pm 7.75 $& $6.66 \pm 2.91$& $-0.09 \pm 1.09 $& $-318 \pm 10\0$\\
\multicolumn{2}{l}{Calibration factor\tablenotemark{e}}&&&&&& 0.1364&3.110&0.329&0.152&0.0172\\
\enddata
\tablenotetext{a}{See Table~\ref{table:regions_table} for region parameters.}
\tablenotetext{b}{The F702W image was taken with WFPC2 in 1994 and 1995, and the calibration has changed since then.  Our calibrations used the \texttt{PHOTFLAM} inverse sensitivity factor converted to janskys via $(F_\nu/{\rm Jy}) = (3.34\times10^4)\cdot (\lambda_{\textrm{peak}}^2) \cdot \texttt{PHOTFLAM} \cdot (\text{counts s}^{-1})$.}
\tablenotetext{c}{Regions 7 and 8 are contaminated by the Einstein ring and at some wavelengths by faint sources and therefore should be considered upper limits.}
\tablenotetext{d}{Background rates are as measured on the science images, which already have had a nominal sky value subtracted.  Units are $10^4$\,counts s$^{-1}$\,pixel$^{-1}$.}
\tablenotetext{e}{Calibration factors are \uJy\,(counts/s)$^{-1}$.}
\tablecomments{Flux densities are in \uJy. 
All tabulated uncertainties are statistical based on the dispersions measured in source-free regions.}
\end{deluxetable*}

\subsection{Modeling X-ray Components with \BAYMAX{}}
\label{subsec:BAYMAX}

Given the extended nature of the X-ray emission, it is difficult to associate X-rays with individual components. For instance, most of the X-ray emission in the Einstein ring region could be associated with the radio lobes, and little X-ray emission is associated with galaxies A or B. The Python tool \BAYMAX{} (Bayesian AnalYsis of Multiple AGNs in X-rays)  calculates the most likely number of components by comparing Bayes factors ($\mathcal{BF}$) of various models. \cite{Foord2019, Foord2020, Foord2021} explained the Bayesian framework behind the tool's calculations. The code simulates the \chandra\ PSF via the ray-trace software {\sc marx}. If a source has multiple observations, the tool can model the PSF (which depends on specifics of each observation such as the detector position) and calculate the likelihoods for each observation. Given the large number of \chandra\ observations for 3C~220.3,  analyzing  each observation individually would have been impractical, and therefore we analyzed only the stacked X-ray image. We used the PSF model of one of the longest observations (Obs ID: 16522) to estimate the likelihoods, but our results are consistent when using the PSF model of any of the  observations. This is not surprising, given that the on-axis \chandra\ PSF has not changed over time.

\begin{figure}[bht!]
\centering
\includegraphics[width=0.45\textwidth,trim={3.5cm 1.2cm 3.1cm 1.2cm},clip]{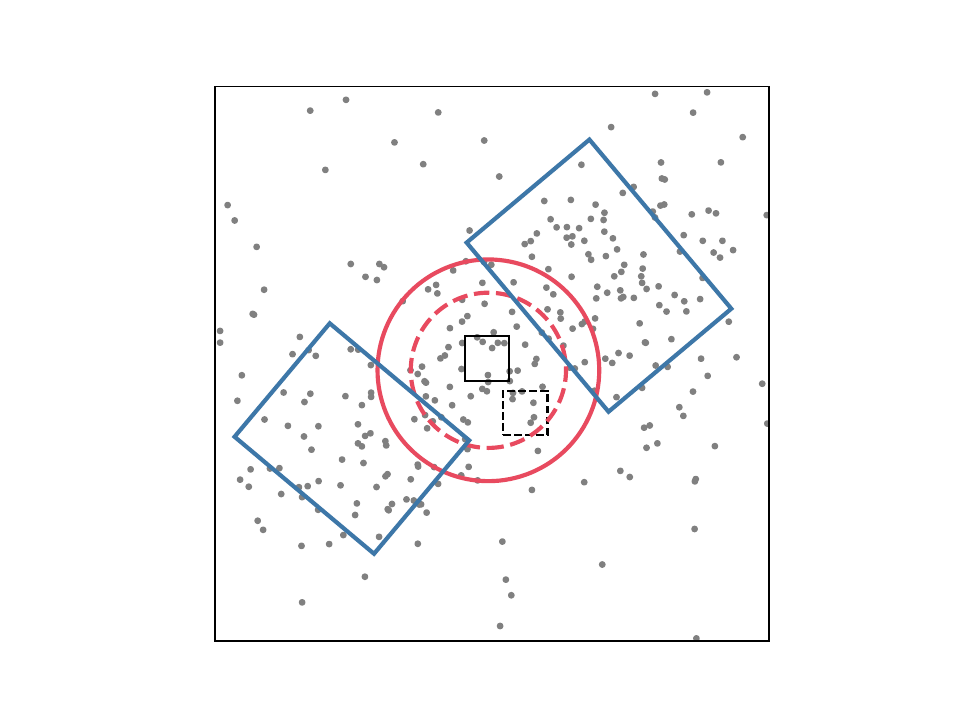}

\caption{Regions used in \BAYMAX{} modeling: galaxy~A (solid black box), galaxy~B (dashed black box), disk (region inside dashed red circle), Einstein ring (annulus bounded by red circles), and radio lobes (outer, rotated blue rectangles). Grey dots represent X-ray photons. Image is about 13\arcsec$\times$13\arcsec\ in size.}
\label{fig:baymax-regions}
\end{figure}

We adapted \BAYMAX{} to evaluate models describing the five components associated with 3C~220.3: galaxy~A, galaxy~B, the NW radio lobe, the SE radio lobe, and the Einstein ring (shown in Figure~\ref{fig:baymax-regions}). All models also included a uniform component to represent the sky background. We added a second set of models with a central circular disk of emission covering the locations of galaxy~A and galaxy~B in lieu of point source emission at either coordinate. Similar to the other regions, photons associated with this component were assumed to be uniformly distributed across the region. We considered models with and without each reasonable combination of components (including sky only, Model~13) as shown in Table~\ref{table:baymax-summary}. For all models, each photon was assumed to originate from either a model component or the sky background, and  counts with energies between 0.5 and 8~keV within a $13\arcsec\times13\arcsec$ box centered on the radio core's coordinates were included.

For the single-point-source models, the probability that a photon observed at detector coordinate $(x,y)$ with energy $E$ is described by the PSF centered at $\mu$ is $P(x,y \mid \mu, E)$, while for the dual point source models the total probability is $P(x, y \mid \mu_{N}, E, f_{n})$. Here $f_{n}$ represents the ratio of counts between the two point sources. The photons associated with the background, Einstein ring, lobe, and extended-galaxy components were assumed to be uniformly distributed across the regions defined in Table~\ref{table:regions_table}. All prior distributions of $\mu$, for both $M_{\mathrm{single}}$ and $M_{\mathrm{dual}}$, were described by continuous uniform distributions. The coordinates of each $\mu$ were defined by the locations of the core, as determined in the VLA radio map shown in Figure~1, and the offset from galaxy~A to~B in the \hst\ images shown in Figure~\ref{fig:new-hst-einstein-rings}. The informative prior distributions for $\mu$ were 1$\arcsec$ across, similar to the regions shown in Figure~\ref{fig:extraction_regions}. The prior distribution of the sky background was described by a uniform distribution (in units of counts arcsec$^{-2}$) centered at a value of 0.11 with a standard deviation of 0.2. The prior distribution for the count ratio, $f_{n}$, was described by a log uniform distribution constrained between $-4$ and 4. 

Across all models, there is strong evidence for an elevated number of counts associated with the radio lobes and with the central area where galaxy~A and galaxy~B sit. However, there is no evidence that galaxy~A and galaxy~B are X-ray point sources; the $\mathcal{BF}$ values favor Model~20, where the central X-rays are modeled as spatially uniform disk. It is difficult to tell whether the Einstein ring component is significant, but there are a few more X-ray counts than expected where the Einstein ring sits, and the $\mathcal{BF}$ is lowest with the ring included.

Since the extended X-ray emission is likely more complicated than our model description, models with many components with arbitrary shapes will be favored. Although \BAYMAX\ favors the Einstein ring component across all comparisons, the best shape to describe the emission in that region is not necessarily a ring. To test the significance of the Einstein ring, we compared Model~20 to a model that replaces the annulus with a rectangular box having uniform emission across the entire region. \BAYMAX\ does not favor this latter, less specific model. These results are not entirely surprising given the spatial distribution of the few X-ray counts in the annulus (Figure~\ref{fig:baymax-regions}), which are insufficient to determine whether an Einstein ring component is statistically significant.

In summary, there is strong evidence for X-ray photons associated with the lobes and with the central area where galaxies~A and~B and the extended galaxy sit. There is no evidence that galaxies~A and~B are X-ray point sources, as the central X-rays are better modeled as spatially uniform emission. It is difficult to be sure whether the Einstein ring is emitting X-rays or not. Deeper X-ray observations would be needed to detect individual components.

\begin{deluxetable}{ccccccc}
\tablecaption{\BAYMAX{} model components and results}
\label{table:baymax-summary}
\tablecolumns{7}
\tablehead{\colhead{Model} & \multicolumn{3}{c}{{Source}} & \colhead{Einstein} & \colhead{Lobes} & \colhead{log}\\[-1.5ex]
\colhead{No.} & \colhead{A} & \colhead{B} & \colhead{disk} & \colhead{Ring}&&\colhead{likelihood\tablenotemark{a}}}
\startdata
1 & \checkmark &  & &  &  & $-$2114.9\\
2 & \checkmark &  & & \checkmark &  & $-$1989.7\\
3 & \checkmark &  & &  & \checkmark & $-$1629.8\\
4 & \checkmark &  & & \checkmark & \checkmark & $-$1571.4\\
5 &  & \checkmark & &  &  & $-$2137.3\\
6 &  & \checkmark & & \checkmark &  & $-$2009.6\\
7 &  & \checkmark & &  & \checkmark & $-$1647.9\\
8 &  & \checkmark & & \checkmark & \checkmark & $-$1589.4\\
9 & \checkmark & \checkmark & &  &  & $-$2098.8\\
10 & \checkmark & \checkmark & & \checkmark &  & $-$1976.5\\
11 & \checkmark & \checkmark & &  & \checkmark & $-$1615.5\\
12 & \checkmark & \checkmark & & \checkmark & \checkmark & $-$1559.9\\
13 &  &  & &  &  & $-$1873.4\\
14 &  &  & & \checkmark &  & $-$1783.8\\
15 &  &  & & & \checkmark & $-$1574.7\\
16 &  &  & & \checkmark & \checkmark & $-$1544.7 \\
17 & & & \checkmark & & & $-$1740.8 \\
18 & & & \checkmark & \checkmark & & $-$1653.5\\
19 & & & \checkmark & & \checkmark & $-$1434.8\\
20 & & & \checkmark & \checkmark & \checkmark & $-$1418.6\\
\enddata
\tablecomments{All models also include a uniform sky background component.}
\tablenotetext{a}{Uncertainties for all log likelihoods are ${\sim} 0.2$}
\end{deluxetable}

\section{Properties of the 3C~220.3 System} \label{sec:discussion}
Having characterized the individual components of 3C 220.3, we can now apply those measurements to examine the properties of the system as a whole.

\begin{figure*}[hbt!]
\includegraphics[width=\textwidth,trim={2.7cm 0cm 2.8cm 1
1.5cm},clip]{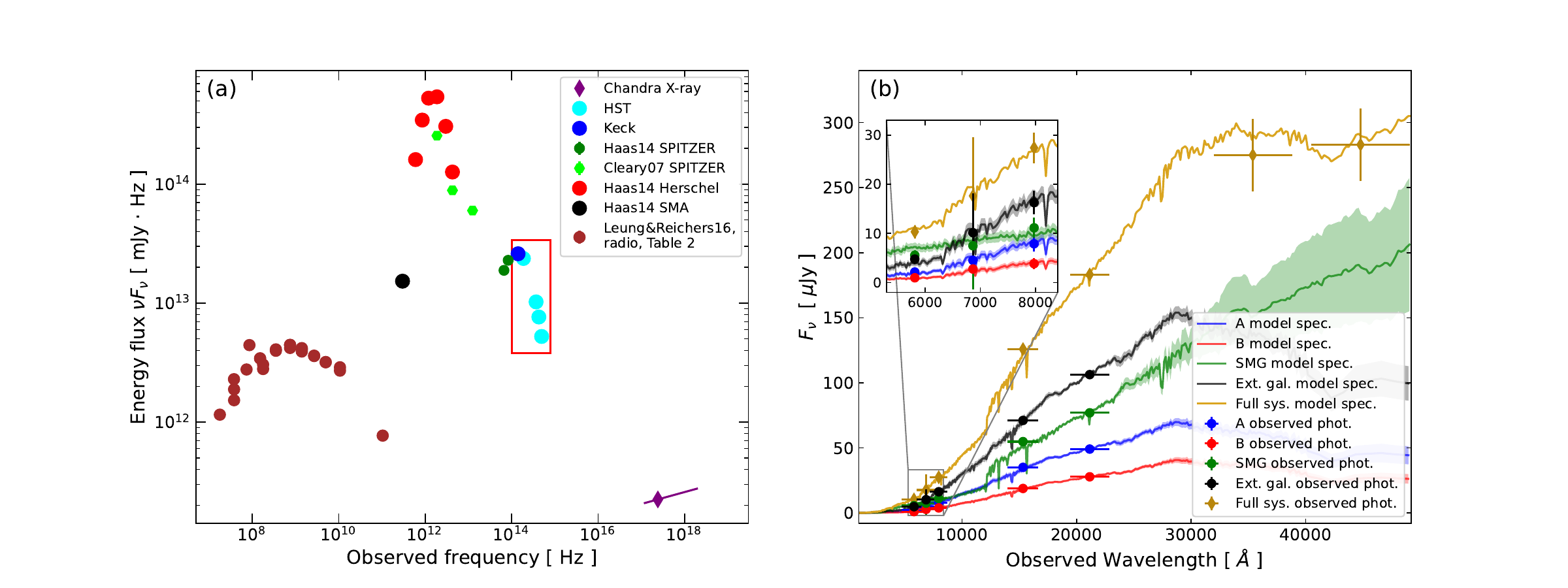}
\caption{(a) Updated SED of the entire 3C~220.3 system.
The optical fluxes include galaxy~A (the AGN host), galaxy~B, the surrounding extended emission,  and the Einstein ring. The X-ray fluxes include the NW and SE lobe regions and were derived from their average flux densities through the Sherpa analysis (Section~\ref{subsec:xray-analysis}). Radio flux densities comprise the radio lobes and core, when detected. Points show observed values from Table \ref{table:sed-values} and \citet[their Table~2]{CARMA_paper} as indicated in the legend. The line through the X-ray point shows the estimated spectral slope, and the red box indicates the wavelength range used in the \Bagpipes\ simulations, shown in panel (b). (b) Visible to NIR flux densities of galaxy~A (blue), galaxy~B (red), extended galaxy (black), and the SMG (green). Stars show measured \hst\ F606W, F702W, F814W, and F160W and Keck $K'$ flux densities. Lines show the simulated, high-resolution \Bagpipes\ spectrum (Section~\ref{subsec:stellar-masses}) for each component. The  full-system sum of both simulated and observed data points, including full-system IRAC 3.6~\micron\ and 4.5~\micron\ flux densities are shown in yellow.}
\label{fig:updated-sed-plots}
\end{figure*}

\label{subsec:sed}
An updated radio-X-ray SED for the full 3C~220.3 system (i.e., sum of all components) is shown in Figure~\ref{fig:updated-sed-plots}a. The corresponding values for the figure are listed in Table 8. Only the visible and near-infrared photometry are able to resolve galaxy A, galaxy B, the extended galaxy, and the Einstein ring, so all flux densities have been summed to determine the full-system optical fluxes. The SMG produces the prominent IR--FIR peak, which would be unusual for a radio galaxy alone \citepalias{Haas_et_al}.

\begin{deluxetable}{cClc}
\tablecaption{SED of entire 3C~220.3 system\tablenotemark{a}}
\label{table:sed-values}
\tablehead{\colhead{Wavelength} & \colhead{Flux Density}  & \colhead{Instrument} & \colhead{Reference} \\[-1.5ex] 
\colhead{(\micron)} & \colhead{(mJy)} & \colhead{} & \colhead{}}
\startdata
0.0012 & 9.26 \times 10^{-7} & Chandra/ACIS-S & 1 \\
0.606 & 0.011 \pm 0.001 & HST/WFC3 & 1\\
0.702 & 0.018 \pm 0.012 & HST/WFPC2 & 2\\
0.814 & 0.028 \pm 0.003 & HST/WFC3 & 1\\
1.6 & 0.127 \pm 0.004 & HST/WFC3 & 1\\
2.124 & 0.184 \pm 0.002 & Keck/NIRC2 & 2 \\
3.6 & 0.275 \pm 0.028 & Spitzer/IRAC & 3 \\
4.5 & 0.283 \pm 0.028 & Spitzer/IRAC & 3 \\
24 & 4.8 \pm 0.5 & Spitzer/MIPS & 5 \\
70 & 30 \pm 5 & Herschel/PACS & 3,4 \\
70 & 20.7 \pm 4.8 & Spitzer/MIPS & 5 \\
70 & 26 \pm 3 & Herschel/PACS & 6 \\
100 & 102 \pm 7 & Herschel/PACS & 3,4 \\
100 & 99 \pm 4 & Herschel/PACS & 6 \\
160 & 289 \pm 9 & Herschel/PACS & 3,4 \\
160 & 136 \pm 31 & Spitzer/MIPS & 5 \\
160 & 259 \pm 11 & Herschel/PACS & 6 \\
250 & 440 \pm 15 & Herschel/SPIRE & 3,4 \\
250 & 452 \pm 9 & Herschel/SPIRE & 6 \\
350 & 403 \pm 20 & Herschel/SPIRE & 3,4 \\
350 & 412 \pm 8 & Herschel/SPIRE & 6 \\
500 & 268 \pm 30 & Herschel/SPIRE & 3,4 \\
500 & 259 \pm 7 & Herschel/SPIRE & 6 \\
1000 & 51 \pm 12 & SMA & 3,4 
\enddata
\tablenotetext{a}{Sum of the flux densities for galaxy~A, galaxy~B, the extended galaxy, and the SMG.}
\tablerefs{1 = this paper, 2 = data from \citetalias{Haas_et_al} remeasured for this paper, 3 = \citetalias{Haas_et_al}, 4 = \citet{CARMA_paper}, 5 = \citet{3C220p3_L_178MHz_Cleary_2007}, 6 = \citet{Westhues_2016}}
\end{deluxetable}

\begin{table*}[hbt!]
\caption{Low-redshift lens properties}
\label{table:summary-table}
\centering
\begin{tabular}{lLLL}
\hline
\hline
 &\rm Galaxy~A &\rm Galaxy~B & \rm Total~A+B\\
\hline
Redshift & 0.6848 \pm 0.0002 & 0.6835 \pm 0.0006 &\no  \\
Stellar Mass $M_*$ & (6.3 \pm 1.5) \times 10^{10}~\Msun & (3.6 \pm 0.9) \times 10^{10}~\Msun & (13.2 \pm 2.7) \times 10^{10}~\Msun\\
Total Mass $M_{\rm tot}$ & (7.99 \pm 0.92) \times 10^{11}~\Msun & (1.10 \pm 0.33) \times 10^{11}~\Msun & (9.09 \pm 0.98) \times 10^{11}~\Msun\\
$M_{\rm tot}({<}0\farcs75)$&(3.90 \pm 0.45) \times 10^{11}~\Msun & (1.23 \pm 0.37) \times 10^{11}~\Msun&\no\\
Dark Matter Fraction & 0.84 \pm 0.04 & 0.71 \pm 0.11 & 0.85 \pm 0.03\\
SFR (A+B only) &\no  & \no & 30^{+14} _{-16}~\Msun\,\text{yr}^{-1}\\
\hline
\end{tabular}
\raggedright
\tablecomments{Uncertainties are 1$\sigma$. Stellar masses for galaxies A and B are from \Bagpipes\ SED fitting within 0\farcs75 radii (Regions 1 and 3 in Table~\ref{table:regions_table}). The stellar mass of ``Total~A+B" uses Region~5b in the \Bagpipes\ SED model. Total masses are from the lens modeling (Section~\ref{subsec:lens-modeling}) and include the full radial extent of each source. $M_{\rm tot}({<}0\farcs75)$ is the lens-model mass within 0\farcs75.
Dark-matter fractions are for $<$0\farcs75 for galaxies A and B and for total masses in the last column.}
\end{table*}

\subsection{Stellar Mass Determinations from Broadband Fluxes}
\label{subsec:stellar-masses}
The luminous masses of galaxies can be estimated using the measured brightness from starlight and luminous gas and dust \citep{1979ARA&A..17..135F_MassLumRatio,Walcher2011_SED_review}. We used the publicly-available \Bagpipes\ code \citep{2018MNRAS.480.4379C} to calculate the stellar masses of galaxy~A (Table~\ref{table:regions_table}, Region No.~1), galaxy~B (Region No.~3), the combined A/B/extended galaxy system (Region No.~5b), and the SMG (Region No.~6). Galaxy~A's fluxes are about double those of galaxy~B for all five visible and NIR bands (Table~\ref{table:HST_Keck_counts_fluxes}). (Only the \hst\ and Keck imaging resolve the different components and are useful in our modeling.) All three galaxies peak in flux density around rest 1.6~\micron. 
The \Bagpipes\ settings included fixed redshifts (\zA\ and \zB\ from Section \ref{subsec:optical-spectra} for galaxies A and B), a delayed exponential star formation history (SFH) model \citep{1986A&A...161...89S_delSFH,2002ApJ...576..135G_delSFH}, \citet{2002MNRAS.336.1188K} stellar initial mass function (IMF), and \citet{dust_Calzetti_2000} dust attenuation law. The resulting stellar masses for galaxies A and B 
are given in Table~\ref{table:summary-table}. With photometry from only five filters, there is a risk of an under-constrained model. A check comes from the observed \spitzer/IRAC flux densities. Even though IRAC did not resolve the components (Figure~\ref{fig:updated-sed-plots}b), the summed  3.6~\micron\ and 4.5~\micron\ model flux densities of galaxy~A, galaxy~B, and the SMG are consistent with the IRAC observations.

\subsection{Lens Modeling}
\label{subsec:lens-modeling}

\begin{figure*}[hbt!]
\centering
\includegraphics[width=\textwidth]{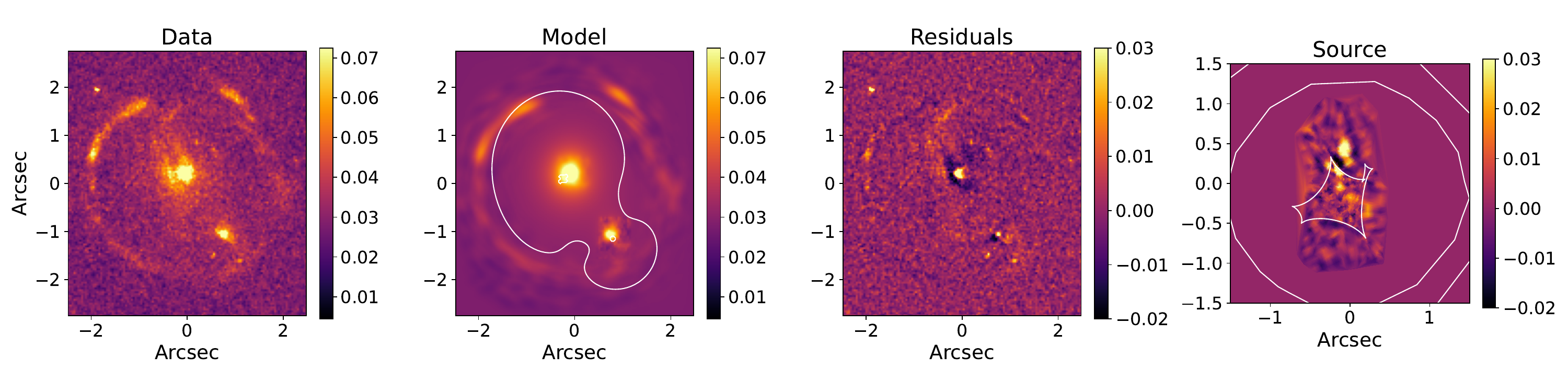}
\caption{Best-fit lens model for the lensed SMG. In order from left to right, panels are: combined \hst\ F814W and F606W image, double-lens model with critical curves overplotted in white, residuals when the model is subtracted from the data, and reconstructed SMG with caustics overplotted in white. The two panels on the right have a different contrast than the two panels on the left.}
\label{fig:lens-model}
\end{figure*}

The high resolution of the new \hst\ imaging enables a more detailed reconstruction of the lensed SMG. The images were modeled with the pixellated Bayesian lens modelling code by \citet{2009_Vegetti_Koopmans_Bayesian_model} as subsequently developed by \citet{2015_Rybak_McKean_Vegetti_lensing}, \citet{2018_Rizzo_Vegetti_3D_lensing}, and \citet{2021_Powell_Vegetti_lensing}. To maximize signal-to-noise, we used the combined \hst\ F814W and F606W image 
(Figure~\ref{fig:lens-model}). The model residuals (second panel from right in Figure~\ref{fig:lens-model}) show an unmodeled point source in the center of the galaxy. This is because we did not model an AGN at the galaxy's center and instead used typical galaxy profiles for galaxies~A and~B.
The resulting best-fit total masses for galaxies~A and~B and their sum are given in Table~\ref{table:summary-table}.

The mass density slope for galaxy~A has a typical isothermal value $\gamma \sim -2$, but galaxy~B is super-isothermal with $\gamma \sim -2.6$ \citep[for comparison, see][]{2006ApJ...649..599K, 2010ApJ...724..511A, 2014MNRAS.437.3670C, 2017MNRAS.464.3742R, 2018MNRAS.475.2403L}. This steeper slope for galaxy~B implies that matter is concentrated more to the center. This could be a result of tidal stripping from a past merger, providing further evidence that galaxies~A and B are interacting or have interacted in the past \citep{2014ApJ...786...89S, 2017MNRAS.464.3742R,2018MNRAS.481.4038L}.

\subsection{Dark Matter Content}
\label{subsec:dark-matter-content}

The stellar masses calculated with our \Bagpipes\ SED fitting (Section~\ref{subsec:stellar-masses}) and the total masses calculated 
from the new lens model imply the dark-matter fractions listed in Table~\ref{table:summary-table}. The fractions are considerably higher than the values calculated by \citetalias{Haas_et_al} ($f_A\sim0.40$ and $f_B\sim0.55$ with uncertainties of $\sim$0.3). Three factors are involved. First, the change from Salpeter IMF to Kroupa IMF decreased the stellar masses by a factor of 0.67 \citep{Madau2014}. Second, \citetalias{Haas_et_al} had only one \hst\ observation (702~nm) in addition to $K'$ and so they assumed an old stellar population. \Bagpipes\ derived younger populations (consistent with the presence of \Oii\ emission) and therefore lower stellar masses, also a factor of $\sim$0.67.  Finally, due to differing extraction regions, our $K'$ flux densities for galaxies A and B  differ from those in H14 by factors of 0.64 and 1.4 respectively, with corresponding differences in stellar mass (Table \ref{table:summary-table}) estimates. For galaxy B the resulting increased mass is outweighed by the other two factors.

Dark matter fractions for galaxies of similar masses and redshifts fall in the range of $0.25 \lesssim f_{\text{DM}} \lesssim 0.75$ \citep{2004ApJ...611..739T,2018MNRAS.481.1950L}, placing galaxy~B at the high end of the distribution and galaxy~A well above typical values. An older stellar population for galaxy~A would decrease its dark-matter fraction, but it is hard to see how it could be as small as 0.75. Although \citetalias{Haas_et_al} suggested that fainter galaxies, i.e., group members, within or obscured by the Einstein ring could increase the stellar mass, the deeper \hst\ images show no evidence of such objects.

\subsection{Radio Lobe X-rays and Magnetic Fields}
\label{subsec:radio-lobe-xrays-bfields}
\label{subsec:radio-lobe-xrays}

Both 3C~220.3 radio lobes show diffuse X-ray emission (Figure~\ref{fig:chandra-native_binned}). X-rays from extended radio structures are generally dominated by inverse-Compton (iC) emission from radio photons upscattered by the high-energy electrons \citep{2011ApJ...729L..12M}. The seed low-energy photons may be those emitted by the synchrotron emission itself (synchrotron self-Compton, SSC) or from the cosmic microwave background (CMB, iC-CMB; \citealt{1998MNRAS.294..615H}). At high redshifts, the CMB photon density is higher, and iC-CMB emission dominates, such as with 3C~270.1 \citep[$z=1.53$;][]{Wilkes_2012_3C270p1}. 3C~220.3 is at medium redshift, so it is useful to determine which process dominates and estimate the magnetic field strength.

The most basic magnetic field strength calculation assumes minimum energy, which is very close to the state of equipartition in which the energy density in the magnetic field matches that in relativistic particles.
Radio-lobe magnetic field strengths determined from the combination of X-ray and radio data are typically a factor of ${\sim}2$--3 below the minimum energy field strength, which can be inferred from the radio synchrotron emission \citep{Worrall_2009_paper} alone.
Table~\ref{table:b-field-summary} gives the minimum-energy field strengths \Bme\ based on the radio flux densities and the lobe radii. These parameters predict less X-ray flux than observed (Table~\ref{table:Chandra_counts_fluxes_HR_lum}), and as usual the required magnetic field is lower than the equipartition value. 

Better estimates of the magnetic field strength come from either of two X-ray emission scenarios: iC-CMB emission only and a combination of iC-CMB and SSC emission.
For iC-CMB alone, the minimum-energy magnetic field can be estimated analytically using Equation~57 of \citet{Worrall_Birkinshaw_2006_Lecture_Notes}, where the constants $C_1$ and $C_2$ are defined via Equations~7 and~56 of that paper. For the internal magnetic field, we used Equation~7 from \citet{Worrall_2009_paper}, simplified for the isotropic, non-relativistic case (as no radio jets are visible). We assumed a ratio of non-radiating heavy particles to radiating electrons $K = 0$ and a volume filling factor $\eta = 1$. We also assumed the range of electron Lorentz factors to be $\gamma_{\text{min}} = 10$ and $\gamma_{\text{max}} = 10^6$ and the radio spectral index $\alpha=0.9$ (particle spectral index $p=2.9$) based on the lobe-dominated radio continuum slope (Figure~\ref{fig:updated-sed-plots}a). Standard CMB properties were used ($\nu_{\text{CMB}} = 282$~GHz, $T_0 = 2.7$~K). The two-dimensional areas of the radio lobes in the plane of the sky were approximated by the elliptical regions listed in Table~\ref{table:regions_table} (lines 7--8). For the magnetic field calculations, we approximated the three-dimensional volumes of the lobes as spheres with the equivalent circular radii of their elliptical counterparts. 
For the pure iC-CMB scenario, the best-fit internal fields for the NW and SE lobes (Table~\ref{table:b-field-summary}) are factors of ${\sim}3.1$ and ${\sim}3.9$ respectively below the equipartition condition.

For the iC-CMB + SSC combination model, 
the internal magnetic fields of the lobes can be calculated numerically with the \texttt{synch} code \citep{1998MNRAS.294..615H}. Table~\ref{table:b-field-summary} gives the calculation inputs and results. In this scenario, the SSC contributions to the X-ray emission are ${\sim}31\%$ and ${\sim}18\%$ for the NW and SE lobes, respectively, and the best-fit internal field strengths are factors of  ${\sim}2.6$ and ${\sim}3.6$ below equipartition. Allowing for the SSC emission raises the internal field estimates to make them closer to the minimum energy values than considering the iC-CMB process alone. 

\begin{table}
\caption{Radio-lobe magnetic field strengths}
\label{table:b-field-summary}
\centering
\begin{tabular}{llcc}
\hline \hline
Quantity & Units & NW Lobe & SE Lobe\\
\hline
\text{Inputs:}\\
~~Equivalent radius\tablenotemark{a}
 &  arcsec& 2.2 & 2.0\\
~~9 GHz flux density & mJy & 118 & 47\\
~~X-ray flux density\tablenotemark{b}
& nJy & 0.270 & 0.226 \\
\hline
\text{Results:}\\
~~\Bme & nT & 8.7 & 7.5 \\
~~$B_{\text{int}}$, iC-CMB-only & nT & $2.8_{-0.6}^{+1.0}$ & $1.9_{-0.4}^{+0.7}$ \\
~~$B_{\text{int}}$, iC-CMB + SSC & nT & $3.4_{-0.7}^{+1.3}$ & $2.1_{-0.4}^{+0.8}$ \\
~~~~SSC X-ray component &  nJy & 0.084 & 0.041 \\
~~~~iC-CMB X-ray compnt. & nJy & 0.186 & 0.185 \\
\hline
\text{$\BmeM/\BintM$ ratios:}\\
~~iC-CMB only & nT & $3.1_{-0.8}^{+0.8}$ & $3.9_{-1.0}^{+1.0}$ \\
~~iC-CMB + SSC & nT & $2.6_{-0.7}^{+0.6}$ & $3.6_{-1.0}^{+0.8}$ \\
\hline
\end{tabular}
\tablenotetext{a}{Equivalent radius is $\sqrt{ab}$ where $a$ and $b$ are the semi-major and semi-minor axes given on lines~7 and~8 of Table~\ref{table:regions_table}.}
\tablenotetext{b}{Same as Table~\ref{table:Chandra_counts_fluxes_HR_lum} last column except for different units.}
\end{table}

\label{subsec:B-field-strengths}

Some of our assumptions could lead to systematic errors in \Bme. First, we assumed that the relativistic particles in the radio lobes are entirely electrons (i.e., $K=0$) rather than some mix including heavy, non-radiating particles such as protons. We also assumed the volume of the radio lobes is completely filled by the magnetic field (i.e., $\eta=1$). However, these assumptions minimize the equipartition fields \citep[Eq.~57]{Worrall_Birkinshaw_2006_Lecture_Notes}, and the disagreement between \Bint\ and \Bme\ would be larger if $K > 0$ or $\eta < 1$.\par

Another possible error in our calculations of the equipartition fields stems from approximating the lobe areas as ellipses and the lobe volumes as spheres. This represents the smallest possible volume with the maximum field strength \citep[Eq.~57]{Worrall_Birkinshaw_2006_Lecture_Notes}. If the true volume were larger, field strengths would be smaller. The power-law slope for the radio and X-ray flux densities has negligible effect on the field strength: changing the slope by 30--50\% changes \Bme/\Bint\ by less than the other uncertainties of the calculations.

Overall, the equipartition fields and internal fields given in Table~\ref{table:b-field-summary} differ by a factor of only ${\sim}$3--4 (well within the range of reported $\BmeM/\BintM$ ratios), and the calculated field strengths are consistent with other 3CR radio galaxies at similar redshifts \citep{2005ApJ...626..733C, Ineson2017}. The SSC emission is significant but not dominant.

\subsection{Extended X-ray emission}
Diffuse X-ray emission is present interior to the radio lobes, spatially overlapping with the extended host-galaxy emission seen in the \hst\ image (Figure~\ref{fig:composite-image}, region No.~5 in Table~\ref{table:regions_table}). While we report an observed X-ray flux for galaxies A (AGN location) and B, our detailed search for structure found no evidence for X-ray point sources (Section~\ref{subsec:BAYMAX}). We therefore conclude that the AGN in this edge-on source is sufficiently obscured to be undetectable in the current data. 

The diffuse X-ray luminosity of $(3.7\pm 1.7) \times 10^{42}$~erg~s$^{-1}$ (Table~\ref{table:Chandra_counts_fluxes_HR_lum}) is comparable to that in the radio lobes, but no extended radio emission is detected in this region. The  3$\sigma$ upper limit $<$1.6~mJy is factors of $\sim$75 and $\sim$30 lower than the radio emission from the NW and SE lobes, respectively. This unusually large X-ray-to-radio flux ratio in comparison with that in the lobes might seem to imply that the X-rays are unlikely to be radio-linked. 

However, that inference is based \citep[as is often done,][]{Worrall_2009_paper}, on an extrapolation of the higher energy radio spectrum observed by the VLA at 9 GHz (Figure~\ref{fig:chandra-native_binned}) to the much lower energy, radio seed photons ($\sim$100--200\,MHz) which up-scatter to provide the X-ray emission. We do not have a reliable observational constraint on these seed photons.

Since the shock front and hot-spots at the head of the jet have moved beyond the host galaxy, the inter-lobe plasma behind it will have cooled down and aged. This likely results in expansion perpendicular to the jet and a reduction in both the particle energies and magnetic field strength in this region. While we cannot constrain the plasma parameters without high-spatial-resolution, low-frequency radio observations, we can estimate whether or not the observed X-ray emission could be generated given our current knowledge. Assuming radio synchrotron emission with the same slope as before ($\alpha = 0.9$), shifted to lower frequencies and constrained not to exceed the total 3C~220.3 source flux at 150~MHz (22.5$\pm$0.08 Jy), we applied a high-energy cut-off due to aging consistent with the  9~GHz radio emission (upper limit 1.6~mJy) in the inter-lobe region. This test model can reproduce the observed inter-lobe, X-ray emission via iC-CMB scattering with a magnetic field ${\sim}5\times$ lower than the equipartition field. This is reasonable given the factors of 3--4 below equipartition reported for the lobes (Table~\ref{table:b-field-summary}). Thus, the current data do not rule out iC-CMB as the primary mechanism generating the extended X-ray emission observed between the radio lobes. Further progress will require a high spatial resolution ($<$0\farcs5, i.e., international baselines) LOFAR observation of 3C~220.3.

An alternative mechanism for the central, diffuse X-ray emission is blends of K$\alpha$ and K$\beta$ lines in  photoionized gas flowing from galactic nuclei, as seen in some local AGN \citep{2011ApJ...736...62W}. This gas also emits the visible \Oiii\,$\lambda 4959,5007$ lines, and a typical ratio of \Oiii\ to soft X-ray emission is $\sim$6--15 \citep{2006A&A...448..499B}. Our long-slit MMT/Binospec spectrum gives a ratio of \Oiii\ line flux to continuum flux within the bandwidth of the \hst\ F814W filter of 1.4\%. Assuming this ratio is constant over the area of diffuse visible  emission,
the \hst-measured flux in the F814W band implies $F(\Oiii) = 1.9 \times 10^{-16}$ erg cm$^{-2}$ s$^{-1}$. This  leads to a ratio of $F(\Oiii)/F_\textrm{X}(\textrm{0.5--2~keV)} \approx 0.1$, $\sim$100~times lower than that seen in local galaxies \citep{2006A&A...448..499B,2009ApJ...704.1195W}.  Therefore, based on the current data, photoionization is  unlikely to explain the diffuse X-ray emission. 

In normal galaxies, the X-ray and 2.2~\micron\ (a proxy for stellar mass) luminosities follow a well-defined scaling relation \citep[][Figure 7.11]{2019cxro.book....7F}. 3C~220.3, with $L(2.2~\micron)\approx1.2\times 10^{10}$~\Lsun, is  over-luminous in X-rays (Table~\ref{table:Chandra_counts_fluxes_HR_lum}) by an order of magnitude relative to the scaling relation. Possible mechanisms for generating this strong, diffuse  X-ray emission include strong star formation or an active galaxy contribution(s). The emission line \Oii\,$\lambda 3727$, a star formation tracer, is unusually strong in 3C~220.3, having a ratio to \Oiii\ $\sim$30~times larger than typical for AGN photoionization, even when ignoring a likely reddening correction. If the ratio of the \Oii\ line to the continuum as measured by Binospec over the \hst\ F606W filter bandwidth is constant over the extended galaxy area, the \hst-measured flux density in the F606W band implies an \Oii\ line flux $F(\Oii) = 1.2\times 10^{-15}$ erg cm$^{-2}$ s$^{-1}$. This line luminosity ${\sim}2.6 \times 10^{42}$ erg s$^{-1}$ implies a SFR of $\sim$17~\Msun~yr$^{-1}$ \citep[][Eq. 4]{2004AJ....127.2002K}, at the low end of the luminous infrared galaxy (LIRG) range. Based on scaling relations \citep[][their Table~3]{Lehmer2016}, this SFR would produce an X-ray luminosity an order of magnitude lower than observed for the extended galaxy, suggesting little contribution to the X-ray emission from star formation  products such as high-mass X-ray binaries. The estimated SFR, while significant, is low relative to the ${\sim}10^{2}~\Msun$~yr$^{-1}$ observed in $\sim$40\%  of the 3CR sources observed by the Herschel satellite \citep{2015A&A...575A..80P}. As such, it is consistent with the large size ($\sim$102 kpc) of the radio jet in a scenario where jet-triggered star formation evolves passively once the jet expands beyond the size of the galaxy \citep{1989MNRAS.239P...1R}.

X-ray emission from the galaxy interstellar medium (ISM) could be the result of shock heating by the passage of the radio jets. This is predicted by models of jet--ISM interaction and has been observed in a number of sources at locations where the ISM lies close to the radio jets, both locally \citep[e.g., NGC~4151;][]{2011ApJ...736...62W} and at high redshift \citep[e.g., 4C~41.17;][]{2000ApJ...540..678B}. In 3C~220.3 this mechanism seems unlikely given that the lobes have progressed beyond the observed physical extent of the galaxy. 

One further possibility is that the diffuse X-ray emission originates in the intra-cluster medium (ICM) of a galaxy group of which 3C~220.3 is a member \citepalias[][Sec~4.3]{Haas_et_al}. 
Typical X-ray luminosities of galaxy groups are ${\sim} 10^{42}$~erg~s$^{-1}$ \citep{Ineson2017}. The  extended emission within 3C~220.3's Einstein ring but not associated with the radio lobes (Table~\ref{table:Chandra_counts_fluxes_HR_lum}, region 5b) has luminosity in this range. 
Additional diffuse X-ray emission to the east and west of the 9~GHz radio lobes (Figure~\ref{fig:composite-image}) has luminosity ${\sim} 3\times 10^{42}$~erg~s$^{-1}$, possibly also consistent with a group scenario, although the emission's elongation along the jet axis would be unusual for a group's ICM.

All in all, the most likely mechanism for the diffuse X-ray emission in 3C~220.3 is iC-CMB from low-frequency radio photons remaining between the radio lobes following the passage of the jet. 
Confirming this interpretation requires deep, high-resolution, low-frequency radio observations  (e.g., with LOFAR) to constrain the presence of radio  emission both within and beyond the radio lobes. 
Absent such confirmation, we cannot rule out diffuse emission from a galaxy group's ICM. While the HST images show several  galaxies near 3C~220.3, follow-up spectroscopy is necessary to confirm whether they are indeed part of a galaxy group.

\section{Conclusions} \label{sec:conclusions}

The strongly lensed SMG in the 3C~220.3 system provides an excellent opportunity to analyze a member of a typically elusive class of objects which occur primarily at high redshifts \citep{Blain_2002_SMGs,Chapman_2005_SMG}. The primary lens, 3C~220.3, is also a powerful radio galaxy, which allows us to study its magnetic field (in the radio lobes), its dark matter content, and its relation to the nearby galaxy~B. This analysis used new, deeper observations from \chandra\ and \hst\ (F606W, F814W, F160W) as well as optical spectroscopy from MMT/Binospec. Our conclusions are:
\begin{enumerate}
    \item \textbf{System Components}---The new \hst\ images have revealed a full Einstein ring of the lensed SMG and extended emission around galaxy~A and possibly galaxy~B. New 20$\times$-deeper \chandra\ data show that diffuse X-ray emission is extended across the entire system, including both radio lobes and around galaxy~A.  The overall spectrum is soft, and there is no evidence for an X-ray point source associated with galaxy~A, consistent with an edge-on AGN being viewed through Compton-thick absorption.
    
    \item \textbf{Redshifts and Emission Lines}---MMT/\discretionary{}{}{}Bino\-spec spectra confirm the redshifts of both galaxy~A (\zA) and the SMG  (\zSMG). Emission lines (\Cii, \Oii, \Hbeta, \Oiii) in galaxy~B's spectrum correspond to \zB, although this could also be attributed to a Doppler shift of ${\sim}$459~\kms\ while galaxies~A and B interact at the same redshift. The extended light around galaxy~A shows \Oii\ emission with complex velocity and spatial structure, but the spectrum is not clearly distinguished from other system components. High-resolution \Oii\ imaging is needed to probe star formation and velocity structure across the system. Combining the observed emission lines with the X-ray analysis of galaxies~A and B suggest that they may both be obscured active galaxies.
    
    \item \textbf{Radio Lobe Magnetic Fields}---Both radio lobes show diffuse X-ray emission consistent with an iC-CMB origin plus a minor contribution from SSC. The X-ray emission from within the radio lobes is best explained by iC-CMB combined with a significant but smaller amount of SSC emission. The derived magnetic fields in the lobes  are factors of ${\sim}$2.6 (NW lobe) and ${\sim}3.6$ (SE lobe) lower than equipartition values, consistent with results for other radio galaxies.
    
    \item \textbf{X-ray Emission from the Extended Galaxy}---Sig\-nifi\-cant diffuse X-ray emission is present between the radio lobes in the area of the diffuse optical emission from galaxy~A.  The most plausible emission mechanism for these X-rays is inverse Compton radiation upscattered from low-frequency radio photons, which can be generated by synchrotron emission from an aging high-energy particle population. The radio flux generated by such a population is unobserved but consistent with the inter-lobe radio upper limit at 9~GHz and the total system radio flux at 150~MHz. High spatial resolution ($\sim$0\farcs5) LOFAR observations are needed to constrain the low-frequency radio emission in the inter-lobe region for confirmation. Photoionization and shock ionization are ruled out as emission mechanisms for these X-rays. However, existing data cannot rule out diffuse X-ray emission from a galaxy group around 3C~220.3.  Spectroscopy of group-galaxy candidates and high-resolution LOFAR data are needed to test the group hypothesis.
    
    \item \textbf{Stellar Masses}---\Bagpipes\ SED fits that include the new \hst\ images give stellar masses of the 3C~220.3 host galaxy~A and its neighbor galaxy~B that are smaller than the masses reported by \citetalias{Haas_et_al}, which were effectively upper limits. Galaxy A (the radio host) has $M_*\approx6\times10^{10}$~\Msun.
    
    \item \textbf{New Lens Model}---An updated lens model using the combined F606W and F814W \hst\ image shows a bright, pointlike residual near the position of galaxy~A. This could represent an AGN seen in rest-frame visible light. Overall, the SMG appears clumpier than \citetalias{Haas_et_al} showed in their model of the $K'$ image. This is consistent with rest-frame visible--UV light being patchier than rest-frame NIR.
    
    \item \textbf{Dark Matter Content}---The ratio of the total mass of the lenses (Table~\ref{table:summary-table})  to the total stellar mass gives a dark matter fraction of the A+B system around 0.85. This is an unusually large value compared to estimates for other radio galaxies.
\end{enumerate}

The 3C~220.3 system presents a rich environment for studying topics from radio lobes to gravitational lensing to a possible galaxy merger. 
The overall appearance, though without complete confidence, is that of an ongoing galaxy merger with the less-massive galaxy tidally stripped and the more-massive galaxy distended by the tidal interaction.  The \Oii\ line's extent both spatially and in velocity indicates star formation over a wide area, likely triggered by the merger. The radio source has size and magnetic field strength typical for its redshift.
Deeper imaging and spectroscopy from \chandra, \hst, and MMT have increased our knowledge of the system, but many intriguing questions remain. What does the \Oii\ emission of the extended galaxy component look like? What is the origin of the diffuse X-ray emission from this component? Are there lensed X-rays from the SMG?

With most of our planet's high-resolution, ground-based optical telescopes located in the Southern Hemisphere, the area near the north celestial pole, where 3C~220.3 resides, is considerably understudied. Telescopes that can observe the region are primarily at low latitudes, such as MMT and Keck, and must contend with high airmasses. Space-based telescopes are the best---if not only---way to conduct certain observations such as the \Oii\ mapping discussed previously. As we look to the coming generation of space-based telescopes, there will be exciting opportunities to turn our sights towards our northern skies and the many extragalactic mysteries that it holds.

\section*{Acknowledgments}
The authors acknowledge with great sadness the death of our long-time collaborator and co-author Professor Mark Birkinshaw in 2023 July. Mark's  contributions to our collaboration and deep knowledge of radio galaxies have always been invaluable to us, and his enthusiasm will continue to be an inspiration. 
We miss him. 

The authors thank Christian Leipski for providing the reduced \hst\ images and Ben Weiner and Sean Moran for their help with the Binospec scheduling and data reduction, respectively.

This research is based on observations made with the NASA/ESA Hubble Space Telescope obtained from the Space Telescope Science Institute, which is operated by the Association of Universities for Research in Astronomy, Inc., under NASA contract NAS 5-–26555. The \hst\ observations of 3C220.3 are associated with program GO-13506 and the work was supported by grant HST-GO-13506.001-A.

Support for this work was provided by the National Aeronautics and Space Administration through Chandra Award Number GO4-15102X issued by the Chandra X-ray Center, which is operated by the Smithsonian Astrophysical Observatory for and on behalf of the National Aeronautics Space Administration under contract NAS8-03060. BJW, JK and MA acknowledge the support of NASA contract: NAS8-03060.
BJW acknowledges the support of the Royal Society and the
Wolfson Foundation while at the University of Bristol, UK. 

S\'{O}H acknowledges support for her research from the Peter Wehinger Graduate Student Fellowship Fund of the Steward Observatory of the University of Arizona.

This research has made use of data obtained from the Chandra Data Archive and the Chandra Source Catalog, and software provided by the Chandra X-ray Center (CXC) in the application packages CIAO, ChIPS, and Sherpa. The scientific results reported in this article are based in part on observations made by the {Chandra X-ray Observatory} using OBSIDs 16081, 16082, 16520, 16521, and~16522.

The National Radio Astronomy Observatory is a facility of the National Science Foundation operated under cooperative agreement by Associated Universities, Inc.

Some of the data presented herein were obtained at the W.\ M.\ Keck Observatory, which is operated as a scientific partnership among the California Institute of Technology, the University of California, and the National Aeronautics and Space Administration. The Observatory was made possible by the generous financial support of the W.\ M.\ Keck Foundation. 

This work is based in part on observations made with the Spitzer Space Telescope, which is operated by the Jet Propulsion Laboratory, California Institute of Technology under a contract with NASA.

This research has made use of the NASA/IPAC Infrared Science Archive, which is funded by the National Aeronautics and Space Administration and operated by the California Institute of Technology. Herschel is an ESA space observatory with science instruments provided by European-led Principal Investigator consortia and with important participation from NASA.

The Submillimeter Array is a joint project between the Smithsonian Astrophysical Observatory and the Academia Sinica Institute of Astronomy and Astrophysics and is funded by the Smithsonian Institution and the Academia Sinica.

The authors wish to recognize and acknowledge the very significant cultural role and reverence that the summit of Maunakea has always had within the indigenous Hawaiian community.  We are most fortunate to have the opportunity to conduct observations from this mountain.

This research made use of SAOImage DS9, developed by Smithsonian Astrophysical Observatory \citep{2000ascl.soft03002S,2003ASPC..295..489J}, Photutils, an Astropy package for detection and photometry of astronomical sources \citep{larry_bradley_2023_7946442} and Regions, an Astropy package for region handling \citep{larry_bradley_2022_7259631}.

The \hst\ and \chandra\ data used in this paper can be accessed at \dataset[10.17909/ghjp-t523]{\doi{10.17909/ghjp-t523}} and \dataset[10.25574/cdc.210]{\doi{10.25574/cdc.210}}, respectively.

\vspace{5mm}
\facilities{HST (WFPC2,WFC3), CXO (ACIS-S), MMT (Binospec), VLA, Keck:II (NIRC2), Keck:I (LRIS), Spitzer (IRAC,MIPS), Herschel (PACS, SPIRE), SMA}

\software{SAOImageDS9 \citep{2000ascl.soft03002S,2003ASPC..295..489J}, Astropy \citep{astropy:2013,astropy:2018,astropy:2022}, Lmfit \citep{2021zndo...4516651N}, CIAO \citep{CIAO_paper}, Sherpa \citep{sherpa:freeman2001a,sherpa:freeman2001b,sherpa:doe2007,sherpa:burke2020}, ChIPS \citep{chips:germain2006}, Bagpipes \citep{2018MNRAS.480.4379C}, BAYMAX \citep{Foord2019, Foord2020, Foord2021}, MARX \citep{2012SPIE.8443E..1AD}, Photutils \citep{larry_bradley_2023_7946442}, Regions \citep{larry_bradley_2022_7259631}, Pyregion, IRAF \citep{1986SPIE..627..733T,1993ASPC...52..173T,1999ascl.soft11002N}}


\bibliography{references}{}
\bibliographystyle{aasjournal}

\end{document}